\documentclass[structabstract]{aa}
\usepackage{txfonts}
\usepackage{graphicx}
\usepackage{natbib}
\usepackage{longtable}
\usepackage{subeqnarray}
\usepackage{cases}
\usepackage{ulem}

\usepackage[colorlinks=true,citecolor=blue]{hyperref}

\begin{document}
\title{Kinetic temperature of massive star-forming molecular clumps measured with formaldehyde}
\subtitle{III. The Orion Molecular Cloud 1}

\author{X. D. Tang\inst{1,2,3}
\and C. Henkel\inst{1,4}
\and K. M. Menten\inst{1}
\and F. Wyrowski\inst{1}
\and N. Brinkmann\inst{1}
\and X. W. Zheng\inst{5}
\and Y. Gong\inst{1,6}
\and Y. X. Lin\inst{1}
\and J. Esimbek\inst{2,3}
\and J. J. Zhou\inst{2,3}
\and Y. Yuan\inst{2,3}
\and D. L. Li\inst{2,3}
\and Y. X. He\inst{2,3}}

\titlerunning{Kinetic temperature in OMC-1}
\authorrunning{X. D. Tang et al.}

\institute{ Max-Planck-Institut f\"{u}r Radioastronomie, Auf dem H\"{u}gel 69, 53121 Bonn, Germany\\
\email{xdtang@mpifr-bonn.mpg.de}
\and Xinjiang Astronomical Observatory, Chinese Academy of Sciences, 830011 Urumqi, PR China
\and Key Laboratory of Radio Astronomy, Chinese Academy of Sciences, 830011 Urumqi, PR China
\and Astronomy Department, King Abdulaziz University, PO Box 80203, 21589 Jeddah, Saudi Arabia
\and Department of Astronomy, Nanjing University, 210093 Nanjing, PR China
\and Purple Mountain Observatory \& Key Laboratory for Radio Astronomy, Chinese Academy of Sciences, 210008 Nanjing, PR China}


\abstract
{We mapped the kinetic temperature structure of the Orion
molecular cloud 1 (OMC-1) with
para-H$_2$CO\,($J_{\rm K_ aK_c}$\,=\,3$_{03}$--2$_{02}$,
3$_{22}$--2$_{21}$, and 3$_{21}$--2$_{20}$)
using the APEX 12\,m telescope. This is compared with the
temperatures derived from the ratio of the NH$_3$\,(2,2)/(1,1) inversion lines
and the dust emission.
Using the RADEX non-LTE model, we derive the gas kinetic
temperature modeling the measured averaged line ratios of
para-H$_2$CO\,3$_{22}$--2$_{21}$/3$_{03}$--2$_{02}$ and
3$_{21}$--2$_{20}$/3$_{03}$--2$_{02}$.
The gas kinetic temperatures derived from the para-H$_2$CO
line ratios are warm, ranging from 30 to >200\,K with an
average of 62\,$\pm$\,2\,K at a spatial density of 10$^5$\,cm$^{-3}$.
These temperatures are higher than those obtained from NH$_3$\,(2,2)/(1,1)
and CH$_3$CCH\,(6--5) in the OMC-1 region.
The gas kinetic temperatures derived from para-H$_2$CO agree
with those obtained from warm dust components measured in the mid infrared (MIR),
which indicates that the para-H$_2$CO\,(3--2) ratios
trace dense and warm gas. The cold dust components
measured in the far infrared (FIR) are consistent with those measured
with NH$_3$\,(2,2)/(1,1) and the CH$_3$CCH\,(6--5) line series.
With dust at MIR wavelengths and para-H$_2$CO\,(3--2) on one side and
dust at FIR wavelengths, NH$_3$\,(2,2)/(1,1), and CH$_3$CCH\,(6--5) on the other,
dust and gas temperatures appear to be equivalent in the dense gas
($n$(H$_2$)\,$\gtrsim$\,10$^{4}$\,cm$^{-3}$) of the OMC-1 region,
but provide a bimodal distribution, one more directly related to
star formation than the other.
The non-thermal velocity dispersions of para-H$_2$CO are
positively correlated with the gas kinetic temperatures
in regions of strong non-thermal motion (Mach number $\gtrsim$\,2.5)
of the OMC-1, implying that the higher
temperature traced by para-H$_2$CO is related to
turbulence on a $\sim$0.06\,pc scale.
Combining the temperature measurements with para-H$_2$CO
and NH$_3$\,(2,2)/(1,1) line ratios, we find direct evidence
for the dense gas along the northern part of the
OMC-1 10\,km\,s$^{-1}$ filament heated by radiation from
the central Orion nebula.}

\maketitle

\section{Introduction}
\label{sect:Introduction}
At a distance of $\approx\,400$\,pc \citep{Menten2007,Kounkel2017}, Orion A and B
are the nearest Giant Molecular Clouds (GMCs) (e.g., \citealt{O'Dell2008}).
Within Orion A, Orion Molecular Cloud 1 (OMC-1) abuts the Orion nebula (M42),
a prominent H\,{\scriptsize II} region  that harbors the Trapezium O-/early
B-type stars and is ionized by one of them, $\theta_1$ C Ori. This star
also excites the Orion Bar, a dense photon-dominated region (PDR), which is a protrusion of
OMC-1 in which the neutral cloud acquires a nearly
edge-on geometry, so that the optically thin PDR emission
is limb-brightened. Star formation in OMC-1 is thought to be taking
place in two locations, in the Kleinmann-Low (KL)  Nebula  and in Orion South.
While evidence exists for low or intermediate mass star formation in
Orion-S \citep{Zapata2004}, the amount of star formation activity in
KL is under debate. In particular, \citet{Zapata2011} have challenged
the view that the famous ``hot core'' located in that region is powered
by a young high mass stellar object.

\begin{figure*}[t]
\centering
\includegraphics[width=1.0\textwidth]{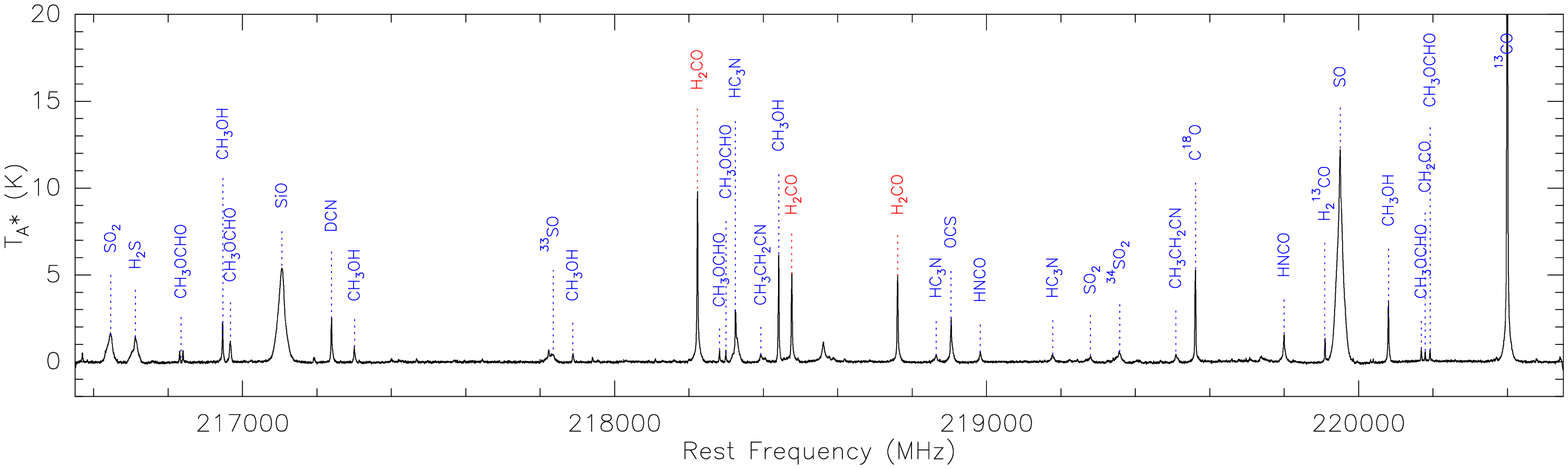}
\caption{A spectrum toward Orion KL obtained with APEX (Atacama Pathfinder EXperiment).
The three H$_2$CO lines discussed in this paper are marked in red.}
\label{fig:KL-spectra}
\end{figure*}

OMC-1 is a unique region for studying
the physical and chemical conditions of molecular clouds.
A large number of molecular line observations have
been performed, such as in
CO (e.g., \citealt{Bally1987,Goldsmith1997,Wilson2005,
Peng2012,Buckle2012,Shimajiri2011,Shimajiri2014,Berne2014}),
CS (e.g., \citealt{Tatematsu1993,Tatematsu1998}),
N$_2$H$^+$ (e.g., \citealt{Tatematsu2008,Melnick2011,Hacar2017}),
NH$_3$ (e.g., \citealt{Batrla1983,Bastien1985,Murata1990,Wiseman1996,Wiseman1998,
Batrla2003,Friesen2017}),
H$_2$CO (e.g., \citealt{Thaddeus1971,Kutner1976,Cohen1981,Batrla1983,Bastien1985,Mangum1990,
Mangum1993b,van der Wiel2009,Leurini2010}),
HC$_3$N (e.g., \citealt{Martin1990,Rodriguez1992,Bergin1996}),
and HCN (e.g., \citealt{Schilke1992,Melnick2011}).
These observations revealed the distribution of the
dense gas within OMC-1 down to a $\sim$0.1\,pc scale. In addition to molecular
line observations, (sub)millimeter continuum observations have revealed remarkable
filamentary structure over the entire Orion A molecular cloud with a length of
$\sim$7\,pc (e.g., \citealt{Lis1998,Johnston1999,Shimajiri2011,Salji2015,Stutz2015,Kainu2017}.

Ammonia (NH$_3$) is frequently
used as the standard molecular cloud thermometer
(e.g., \citealt{Ho1983,Walmsley1983,Danby1988,Mangum2013b}).
However, it has shown a large spread in fractional abundance (e.g., 10$^{-5}$
in dense molecular "hot cores" around newly formed
massive stars, \citealt{Mauersberger1987}; 10$^{-8}$ in
dark clouds, \citealt{Benson1983,Chira2013}; $\sim$10$^{-10}$
in the Orion Bar PDR, \citealt{Batrla2003}) and is extremely affected
by a high UV flux. In contrast, formaldehyde (H$_2$CO) is a more ubiquitous molecule
in the interstellar medium \citep{Downes1980,Bieging1982,Cohen1983,Baan1986,Baan1990,
Baan1993,Henkel1991,Zylka1992,Mangum2008,Mangum2013a,Zhang2012,Ao2013,Tang2013,Ginsburg2015,
Ginsburg2016,Guo2016}. Unlike ammonia, the fractional abundance of H$_2$CO is stable
at various stages of star formation \citep{Mangum1990,Caselli1993,Mangum1993b,
Johnstone2003,Gerner2014,Tang2017a,Tang2017b,Tang2017c}.
OMC-1 has been a valuable target for measuring lines of H$_2$CO because of its
high densities and temperatures
(e.g., \citealt{Mangum1990,Mangum1993b,Bergin1994,Bergin1996,Peng2012,Gong2015b,Kauffmann2017})
and subsequently large surface brightnesses (e.g., \citealt{Lis1998,Johnston1999,Megeath2012,Lombardi2014}).
Previous observations show that H$_2$CO has a spatially
extensive distribution in the OMC-1 region including Orion KL,
Orion south, the Orion Bar, and the northern part of the OMC-1 region
(e.g., \citealt{Batrla1983,Bastien1985,Mangum1990,Mangum1993b,van der Wiel2009,Leurini2010}).

Formaldehyde is a reliable probe to trace physical conditions of molecular clouds
\citep{Henkel1980,Henkel1983,Mangum1993a,Muhle2007,Ginsburg2011}.
It has a rich variety of millimeter and submillimeter transitions.
Particularly useful are the three transitions of para-H$_2$CO
($J_{\rm K_aK_c}$\,=\,3$_{03}$--2$_{02}$, 3$_{22}$--2$_{21}$, and
3$_{21}$--2$_{20}$), which can be measured simultaneously at
$\sim$218\,GHz with a bandwidth of 1\,GHz.
These transitions of para-H$_2$CO may trace denser regions
than NH$_3$ (1,1)/(2,2) (e.g., \citealt{Ott2014,Ginsburg2016}).
Since the relative populations of the $K_{\rm a}$ ladders of
H$_2$CO are governed by collisions, line ratios involving
different $K_{\rm a}$ ladders are good tracers of the kinetic
temperature \citep{Mangum1993a,Muhle2007}.
Therefore the line strength ratios of para-H$_2$CO,
3$_{22}$--2$_{21}$/3$_{03}$--2$_{02}$
and 3$_{21}$--2$_{20}$/3$_{03}$--2$_{02}$, provide a sensitive
thermometer, possibly the best of the very few which are
available for the analysis of dense molecular gas.
These H$_2$CO line ratios have been
used to measure physical parameters in our
Galactic center clouds \citep{Qin2008,Ao2013,Johnston2014,
Ginsburg2016,Immer2016,Lu2017}, star formation regions
\citep{Mangum1993a,Hurt1996,Mitchell2001,Watanabe2008,Lindberg2015,Tang2017a,Tang2017c},
as well as in external galaxies \citep{Muhle2007,Tang2017b}.

Many observations have been performed to reveal the temperatures of gas and dust
in the Orion region (e.g., \citealt{Downes1981,Churchwell1983,Bergin1994,Wiseman1998,
Mookerjea2000,Vaillancourt2002,Batrla2003,Megeath2012,Peng2012,
Lombardi2014,Goicoechea2015,Nishimura2015,Salgado2016,Friesen2017}).
However, critical links between H$_2$CO and other measurements of
gas temperatures as well as dust temperatures are still unclear.
In this paper, we aim to map the kinetic temperature structure
of the OMC-1 region with three transitions of
para-H$_2$CO\,($J_{\rm K_aK_c}$\,=\,3$_{03}$--2$_{02}$, 3$_{22}$--2$_{21}$,
and 3$_{21}$--2$_{20}$).
In Sections \ref{sect:Observation} and \ref{sect:Results},
we introduce our observation of the para-H$_2$CO triplet,
data reduction, and results. We discuss the resulting
kinetic temperatures derived from
para-H$_2$CO in Section \ref{sect:discussion}.
Our main conclusions are summarized in Section \ref{sect:summary}.
This paper is part of the "Kinetic temperature of massive
star-forming molecular clumps measured with formaldehyde" series of
studies exploring H$_2$CO as a probe of gas conditions in a variety of sources.

\begin{figure*}[t]
\centering
\includegraphics[width=0.53\textwidth,angle=-90]{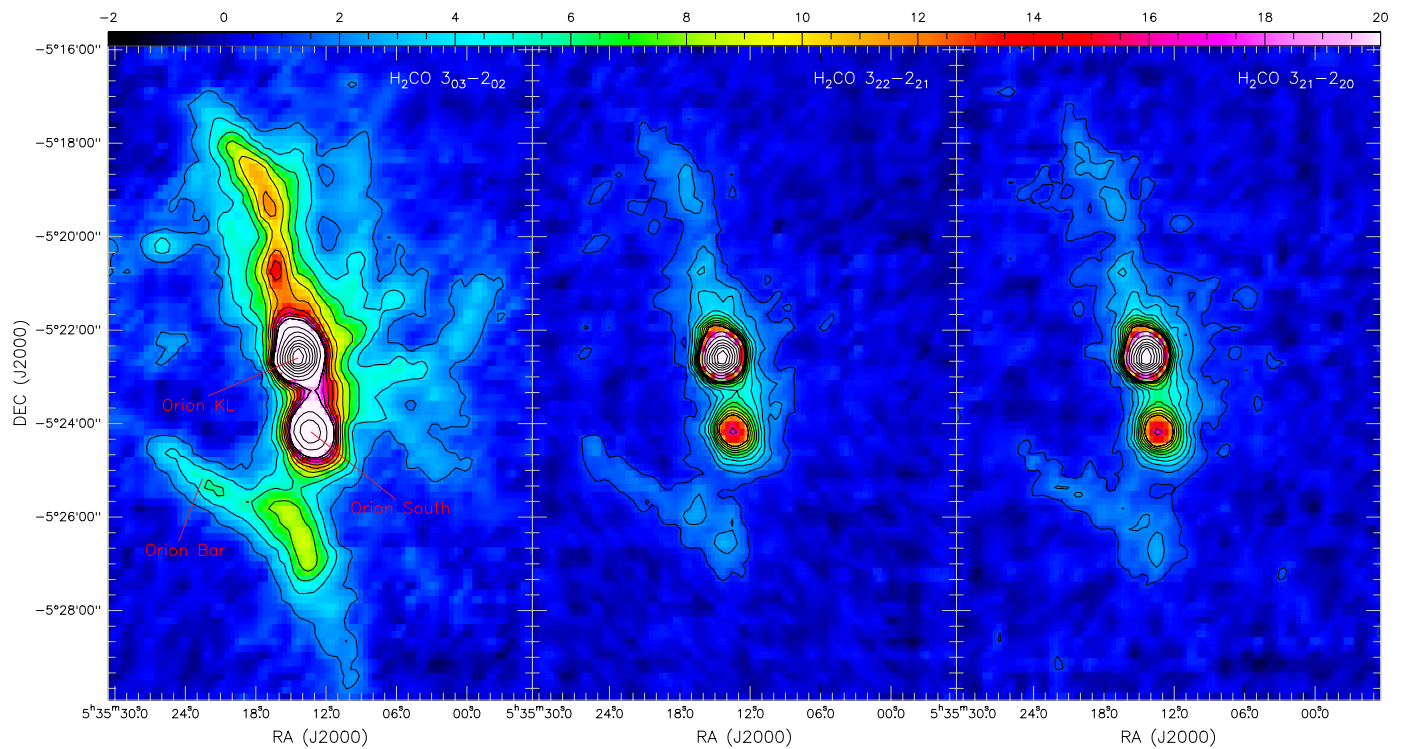}
\caption{Intensity maps ($T$$_{\rm A}^*$ scale; color bar in units of
K\,km\,s$^{-1}$) of para-H$_2$CO\,3$_{03}$--2$_{02}$ (left),
3$_{22}$--2$_{21}$ (middle), and 3$_{21}$--2$_{20}$ (right)
integrated from $V_{\rm LSR}$\,=\,4 to 14\,km\,s$^{-1}$ of the OMC-1 cloud.
Contour levels are from 1.9 to 19\,K\,km\,s$^{-1}$ with steps of
1.9\,K\,km\,s$^{-1}$ and from 19 to 95\,K\,km\,s$^{-1}$ with steps
of 9.5\,K\,km\,s$^{-1}$ for para-H$_2$CO\,3$_{03}$--2$_{02}$,
and from 1.05 to 10.5\,K\,km\,s$^{-1}$ with steps of 1.05\,K\,km\,s$^{-1}$
and from 10.5 to 52.5\,K\,km\,s$^{-1}$ with steps of 5.25\,K\,km\,s$^{-1}$
for para-H$_2$CO\,3$_{22}$--2$_{21}$ and 3$_{21}$--2$_{20}$.}
\label{fig:H2CO-maps}
\end{figure*}

\section{Observations and data reduction}
\label{sect:Observation}
Our observations were carried out on 2015 October with
the APEX\footnote{\tiny
This publication is based on data acquired with the
Atacama Pathfinder EXperiment (APEX). APEX is a
collaboration between the Max-Planck-Institut f\"{u}r
Radioastronomie, the European Southern Observatory,
and the Onsala Space Observatory.} 12\,m telescope located
on Chajnantor (Chile) using the APEX-1 (SHeFI) receiver.
The beam size is $\sim$\,28.5$''$ ($\sim$0.06\,pc at $\sim$400\,pc distance;
see \citealt{Menten2007,Kounkel2017}).
The main beam efficiency and the forward efficiency were 0.75
and 0.97, respectively. The para-H$_2$CO
$J_{\rm K_aK_c}$\,=\,3$_{03}$--2$_{02}$, 3$_{22}$--2$_{21}$, and 3$_{21}$--2$_{20}$
transitions have rest frequencies of 218.222, 218.475, and 218.760\,GHz,
respectively, which are measured simultaneously by employing the
eXtended bandwidth Fast Fourier Transform Spectrometer (XFFTS)
backend for two spectral windows of 2.5\,GHz bandwidth.
The frequency range, consisting of two spectral windows,
covered 216.5--219.0\,GHz and 218.0--220.5\,GHz with the central frequency being set to
218.550\,GHz. 32768 spectral channels were used in each window.
This provides a velocity resolution of $\sim$0.1\,km\,s$^{-1}$.
The on-the-fly observing mode was used to measure eight
4.5$\times$3.5\,arcmin$^2$ maps with steps of 9$''$ in both
right ascension and declination. The surveyed area of the OMC-1
is $\sim$9$\times$14\,arcmin$^2$ ($\sim$1.1$\times$1.7\,pc$^2$),
centered on $\alpha_{2000}$\,=\,05:35:12.50, $\delta_{2000}$\,=\,--05:22:55.0.
The observed spectrum toward Orion KL
is shown in Figure \ref{fig:KL-spectra}.

Data reduction for spectral lines and maps was performed with
GILDAS\footnote{\tiny http://www.iram.fr/IRAMFR/GILDAS}.
The spectra were resampled in steps of $\sim$14.2$''$. To enhance
signal to noise ratios (S/Ns) in individual channels, we smoothed contiguous
channels to a velocity resolution $\sim$0.4\,km\,s$^{-1}$.
A typical rms noise level (1$\sigma$)
is $\sim$0.13\,K ($T_{\rm mb}$ scale) for
a 0.4\,km\,s$^{-1}$ wide channel. In total we acquired 2472 positions,
corresponding to 2472 spectra for each transition.
We fit all spectra with Gaussian profiles.
Nearly 50\% of all positions were detected in para-H$_2$CO\,3$_{03}$--2$_{02}$
and $\sim$16\% were also detected in para-H$_2$CO\,3$_{22}$--2$_{21}$ and
3$_{21}$--2$_{20}$, respectively, with signal-to-noise ratios of $\gtrsim$3$\sigma$.

\section{Results}
\label{sect:Results}
\subsection{Distribution of H$_2$CO}
The intensity distributions of the three transitions of para-H$_2$CO integrated
from 4 to 14\,km\,s$^{-1}$ in the OMC-1 are shown in Figure \ref{fig:H2CO-maps}.
The para-H$_2$CO\,3$_{03}$--2$_{02}$ line shows an extended
distribution and clearly traces the dense molecular structure,
e.g., Orion KL, Orion South, the Orion Bar, and molecular fingers, which
confirms previous observational results of the dense gas distribution traced by H$_2$CO
(e.g., \citealt{Thaddeus1971,Kutner1976,Batrla1983,Bastien1985,Mangum1990,
van der Wiel2009,Leurini2010}).
It is also consistent with
previous observational results probed with other dense gas tracers of
e.g., C$^{18}$O, CS, CN, HCN, HNC, HCO$^+$, HC$_3$N, CH$_3$CCH
\citep{Martin1990,Rodriguez1992,Schilke1992,Tatematsu1993,Tatematsu1998,
Bergin1994,Bergin1996,Goldsmith1997,Ungerechts1997,Buckle2012,
Peng2012,Shimajiri2014}.
Para-H$_2$CO\,3$_{22}$--2$_{21}$ and 3$_{21}$--2$_{20}$
are only detected in the densest regions of the OMC-1,
and show less extended distributions than para-H$_2$CO\,3$_{03}$--2$_{02}$.

\begin{figure*}[t]
\centering
\includegraphics[width=0.88\textwidth]{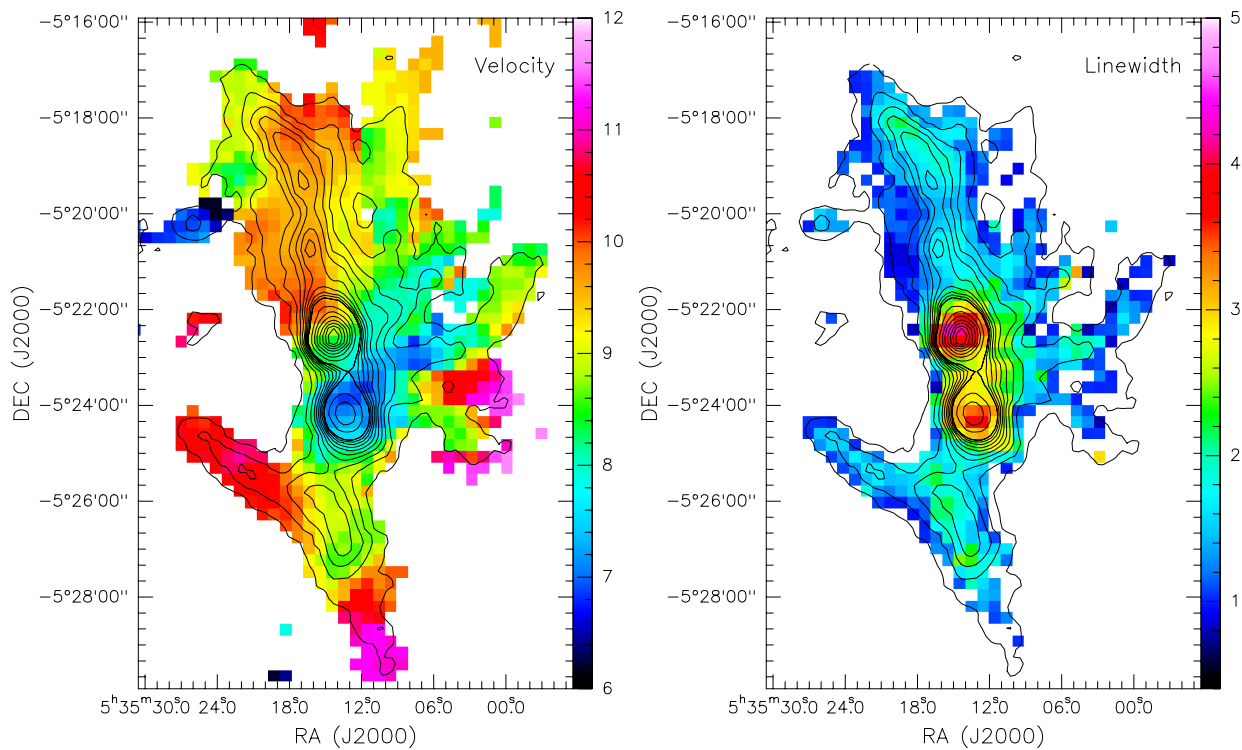}
\caption{The para-H$_2$CO\,3$_{03}$--2$_{02}$ intensity-weighted
mean velocity map (moment 1, left) and velocity dispersion map (moment 2, right).
The unit of each colour bar is km\,s$^{-1}$. Contours are
para-H$_2$CO\,3$_{03}$--2$_{02}$ integrated intensities
(same as in Figure \ref{fig:H2CO-maps}).}
\label{fig:H2CO-velocity-width-distributions}
\end{figure*}

The para-H$_2$CO\,3$_{03}$--2$_{02}$ intensity-weighted mean
velocity (moment 1), and velocity dispersion (moment 2) maps are
shown in Figure\,\ref{fig:H2CO-velocity-width-distributions},
revealing complex structure. Figure\,\ref{fig:H2CO-Channel-map}
shows the corresponding channel maps. The northeastern region and the
star forming regions Orion KL and Orion South with their outflows are
prominent at low velocities ($\sim$6--9\,km\,s$^{-1}$),
while the Orion Bar and the northern region become visible at higher
velocities ($\sim$9--11\,km\,s$^{-1}$).
The velocity dispersion of para-H$_2$CO\,(3$_{03}$--2$_{02}$) ranges from
0.7 to 6\,km\,s$^{-1}$ with an average of $\sim$2\,km\,s$^{-1}$.
Only the star forming regions Orion KL and Orion South are
prominent at high velocity dispersions ($>$3\,km\,s$^{-1}$).
In other dense regions of the OMC-1, the velocity dispersions become
$\sim$1.5--3\,km\,s$^{-1}$. The outer parts of the OMC-1
show a lower range of velocity dispersion ($<$1.5\,km\,s$^{-1}$).

The distribution of para-H$_2$CO\,(3$_{03}$--2$_{02}$) is similar to
the spatial distribution of the NH$_3$\,(1,1) and (2,2) emission
mapped with the Green Bank Telescope (GBT)
(beam size $\sim$30$''$; \citealt{Friesen2017}) (see Figure \ref{fig:NH3-850um})
and all prominent features identified in the NH$_3$\,(1,1) and (2,2)
lines are detected in para-H$_2$CO\,(3$_{03}$--2$_{02}$)
at a scale of $\sim$0.06\,pc in the OMC-1.
The similar distributions of H$_2$CO\,(3$_{03}$--2$_{02}$)
and NH$_3$\,(1,1) and (2,2) in the OMC-1 region are consistent
with previous observational results in the Galactic Central
Molecular Zone (CMZ) on a $\sim$0.1\,pc scale \citep{Lu2017}.
The strongest H$_2$CO\,(3$_{03}$--2$_{02}$) emission
associates with the massive star formation regions
Orion KL and Orion South, which is consistent with the NH$_3$\,(2,2) emission.
However, the strongest NH$_3$\,(1,1) emission is found toward Orion KL
and the northern clumps, which is slightly different from what
we see in H$_2$CO\,(3$_{03}$--2$_{02}$). The  agreement between
the distributions of para-H$_2$CO\,(3$_{03}$--2$_{02}$)
and NH$_3$\,(2,2) is better than that between para-H$_2$CO\,(3$_{03}$--2$_{02}$)
and NH$_3$\,(1,1) (see Figure \ref{fig:NH3-850um}).
The probable reason is that
NH$_3$\,(1,1) may sample a more extended and lower density gas.

The para-H$_2$CO\,(3$_{03}$--2$_{02}$)
integrated intensity distribution agrees remarkably well
with the 450 and 850\,$\mu$m dust emission \citep{Johnston1999} in the OMC-1 region,
including the dense molecular structure and
dust emission peaks (see Figure \ref{fig:NH3-850um}).
This confirms previous observational results in massive
star-forming clumps at various evolutionary stages \citep{Tang2017a,Tang2017c}.

\subsection{Kinetic temperature}
\label{sect:Kinetic-temperature}
As discussed in Section \ref{sect:Introduction}, the relative intensity
ratio of H$_2$CO lines involving different $K_{\rm a}$ ladders yields estimates
of the kinetic temperature of the gas \citep{Mangum1993a}.
The para-H$_2$CO\,3$_{22}$--2$_{21}$ and 3$_{21}$--2$_{20}$ transitions
have similar upper state energies above the ground state, $E_{\rm u}$\,$\simeq$\,68\,K,
similar spatial distributions (see Figure \ref{fig:H2CO-maps}),
similar line profiles (brightness temperature, linewidth,
and velocity in our observations; see Figure\,\ref{fig:H2CO322-321};
also see \citealt{Tang2017a,Tang2017b,Tang2017c}),
and are often detected together in molecular clouds
(e.g., \citealt{Bergman2011,Wang2012,Lindberg2012,Ao2013,Immer2014,
Trevino2014,Ginsburg2016,Ginsburg2017,Tang2017a,Tang2017b,Tang2017c,Lu2017}).
The CH$_3$OH\,4$_{22}$--3$_{12}$ transition at 218.440\,GHz is well
separated from the para-H$_2$CO\,(3$_{22}$--2$_{21}$) transition
in the OMC-1 region (see Figures\,\ref{fig:KL-spectra} and \ref{fig:H2CO322-321}).
Para-H$_2$CO\,3$_{22}$--2$_{21}$/3$_{03}$--2$_{02}$ and
3$_{21}$--2$_{20}$/3$_{03}$--2$_{02}$ ratios are good thermometers
to determine kinetic temperature and show a similar behavior
to kinetic temperature and spatial density at high density
$n$(H$_2$)\,$\gtrsim$\,10$^5$\,cm$^{-3}$ \citep{Lindberg2015,Tang2017a},
so in this work we use the averaged ratio
0.5$\times$[(3$_{22}$--2$_{21}$\,+\,3$_{21}$--2$_{20})$/3$_{03}$--2$_{02}$]
between para-H$_2$CO\,3$_{22}$--2$_{21}$/3$_{03}$--2$_{02}$ and
3$_{21}$--2$_{20}$/3$_{03}$--2$_{02}$ to determine gas kinetic temperatures.

\begin{figure}[t]
\vspace*{0.2mm}
\begin{center}
\includegraphics[width=0.49\textwidth]{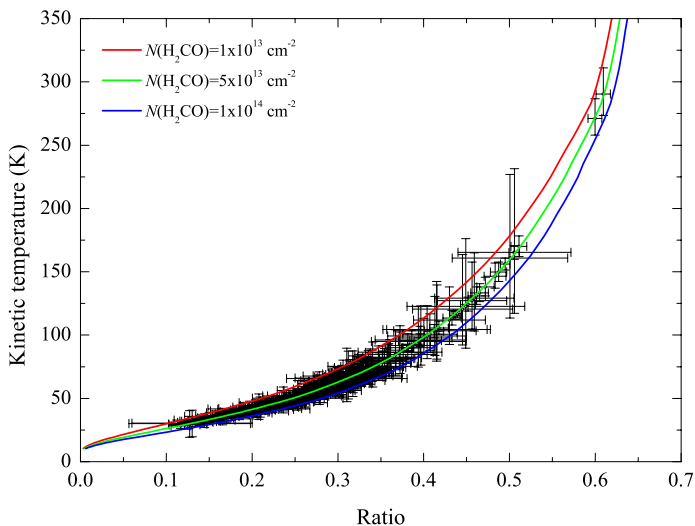}
\end{center}
\caption{RADEX non-LTE modeling of the relation between the
kinetic temperature and the average ratio of
para-H$_2$CO\,3$_{22}$--2$_{21}$/3$_{03}$--2$_{02}$ and
3$_{21}$--2$_{20}$/3$_{03}$--2$_{02}$ with an assumed density of
$n$(H$_2$)\,=\,10$^{5}$\,cm$^{-3}$, an averaged linewidth of 2.0\,km\,s$^{-1}$,
and column density of $N$(para-H$_2$CO)\,=\,1$\times$10$^{13}$ (red line),
5$\times$10$^{13}$ (green line), and 1$\times$10$^{14}$\,cm$^{-2}$ (blue line).
The black points are derived from our observed H$_2$CO line ratios for a column density
$N$(para-H$_2$CO)\,=\,5$\times$10$^{13}$\,cm$^{-2}$, the average
value for the OMC-1 (see Section \ref{sect:Kinetic-temperature}). The temperature
uncertainties are obtained from observed para-H$_2$CO line ratio errors.}
\label{fig:Tk-H2CO-ratios}
\end{figure}

Using the RADEX\footnote{\tiny http://var.sron.nl/radex/radex.php}
non-LTE model \citep{van der Tak2007} with collision
rates from \cite{Wiesenfeld2013}, we modeled the relation between
the gas kinetic temperature and the measured average of
para-H$_2$CO\,0.5$\times$[(3$_{22}$--2$_{21}$\,+\,3$_{21}$--2$_{20})$/3$_{03}$--2$_{02}$]
ratios, adopting an average measured linewidth of $\sim$2.0\,km\,s$^{-1}$
(temperature weakly dependent on the linewidth; see \citealt{Nagy2012,Immer2016})
and column densities $N$(para-H$_2$CO)\,=\,1$\times$10$^{13}$, 5$\times$10$^{13}$,
and 1$\times$10$^{14}$\,cm$^{-2}$ in Figure \ref{fig:Tk-H2CO-ratios}.
Previous observations toward the Galactic CMZ clouds and dense
massive star-forming clumps show that the opacities of para-H$_2$CO\,(3--2)
lines weakly influence the
measurements of gas kinetic temperature \citep{Ginsburg2016,Immer2016,Tang2017c},
so here we assume that the para-H$_2$CO\,(3--2) lines are optically thin in the OMC-1 region.
The spatial densities measured with H$_2$CO and HC$_3$N in the OMC-1
region are $n$(H$_2$)\,$\sim$\,10$^{5}$--10$^{6}$\,cm$^{-3}$ \citep{Mangum1993b,Bergin1996}.
Previous observations show that para-H$_2$CO\,(3--2) is sensitive
to gas at density of 10$^{5}$\,cm$^{-3}$ \citep{Ginsburg2016,Immer2016,Tang2017b}.
Therefore, we adopt 10$^{5}$\,cm$^{-3}$ as an average gas spatial density in the OMC-1 region.
The average column density $N$(para-H$_2$CO) obtained from the para-H$_2$CO\,(3$_{03}$--2$_{02}$)
averaged brightness temperatures of the entire OMC-1 region at density 10$^{5}$\,cm$^{-3}$
using the method of \cite{Tang2017a} is $\sim$5$\times$10$^{13}$\,cm$^{-2}$.
The temperatures derived from different column densities at density 10$^{5}$\,cm$^{-3}$
would only slightly change (see Figure \ref{fig:Tk-H2CO-ratios}).
Generally, higher line ratios of para-H$_2$CO indicate higher kinetic
temperatures. Therefore, the ratio maps can be used as a proxy
for relative kinetic temperature.
We use the relation between kinetic temperature and para-H$_2$CO line ratios
at spatial density 10$^{5}$\,cm$^{-3}$ and column density
5$\times$10$^{13}$\,cm$^{-2}$ (shown in Figure \ref{fig:Ratio-Tk-map}) to
convert ratio maps to temperature maps in Figure \ref{fig:Ratio-Tk-map}.

An averaged para-H$_2$CO\,(3--2) line
ratio map is shown in Figure \ref{fig:Ratio-Tk-map}.
The line ratios are calculated by velocity-integrated intensities where
the para-H$_2$CO\,3$_{22}$--2$_{21}$ and/or 3$_{21}$--2$_{20}$ lines are
detected with S/N\,$\gtrsim$\,3$\sigma$. Para-H$_2$CO ratios range from 0.12 to 0.61,
with an average of 0.28\,$\pm$\,0.01 (errors given here and elsewhere
are standard deviations of the mean). The lowest ratios ($<$0.25)
occur in the Orion north(west) and
the ridge between Orion South and the Orion Bar. In Orion South, in the north(east),
and in the dense clumps in the Orion Bar the ratios range from 0.25 to 0.4.
High ratios ($>$0.4) are also found
in the northeastern region, Orion KL, and the Orion Bar.
Several locations near the hot core, in the
northeastern region, and in the Orion Bar around the H\,{\scriptsize II} region
show the highest ratios ($>$0.45).

The gas kinetic temperatures derived from the
para-H$_2$CO line ratios are warm, ranging from 30 to $>$200\,K
with an average of 62\,$\pm$\,2\,K at density $n$(H$_2$)\,=\,10$^5$\,cm$^{-3}$,
which agrees with the results measured with H$_2$CO in other star-forming
regions \citep{Mangum1993a,Hurt1996,Mangum1999,Watanabe2008,Nagy2012}
and Galactic center clouds \citep{Ao2013,Ginsburg2016,Immer2016,Lu2017}.
The kinetic temperatures in the dense gas around the H\,{\scriptsize II}
region and in the
northeastern part of the OMC-1 10\,km\,s$^{-1}$ filament are high
($>$50\,K; see Figure \ref{fig:Ratio-Tk-map} and Table\,\ref{table:Temperatures}).
Typical kinetic temperatures are $\sim$120--290\,K in Orion KL, $\sim$74\,K
in Orion
South, $\sim$73\,K in dense clumps of the Orion Bar, $\sim$100--130\,K
at the edge (north and south) of the Orion Bar, $\sim$40\,K in Orion
north, and $\sim$70--160\,K in the Orion northeastern region.

\section{Discussion}
\label{sect:discussion}
\subsection{Comparison of temperatures derived from H$_2$CO and other gas tracers}
\label{Sec:Tk-H2CO-others}
\begin{figure*}[t]
\centering
\includegraphics[width=0.98\textwidth]{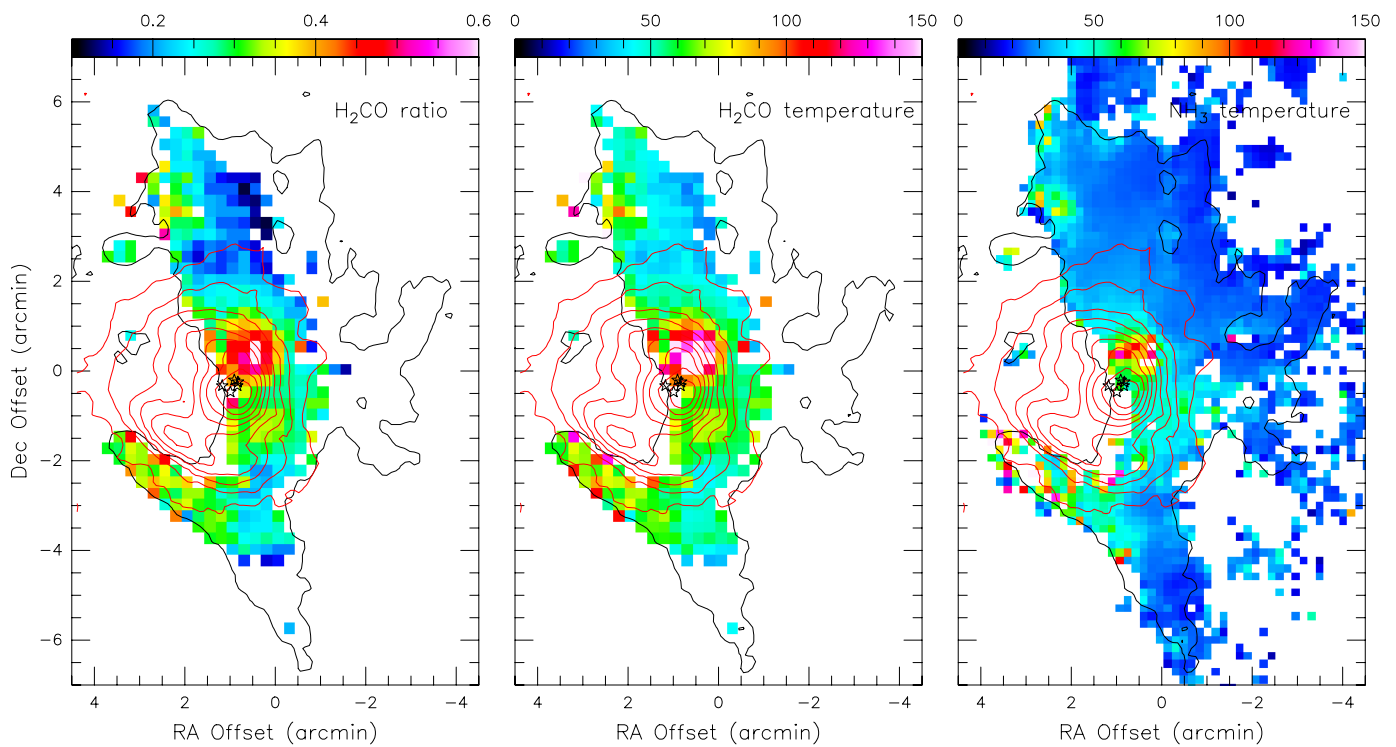}
\caption{Left: The averaged (see Section \ref{sect:Kinetic-temperature})
velocity-integrated intensity ratio map of
para-H$_2$CO 0.5$\times$[(3$_{22}$--2$_{21}$\,+\,3$_{21}$--2$_{20})$/3$_{03}$--2$_{02}$]
in the OMC-1. Middle: The kinetic temperatures derived from the
para-H$_2$CO\,(3--2) line ratios. Right: The kinetic temperatures derived from NH$_3$\,(2,2)/(1,1)
ratios observed with the GBT (beam size $\sim$30$''$; \citealt{Friesen2017}).
Black contours show the contour of integrated intensity of para-H$_2$CO\,3$_{03}$--2$_{02}$
at 1.9\,K\,km\,s$^{-1}$ (see Figure \ref{fig:H2CO-maps}).
Red contours show the velocity-integrated intensity
($V_{\rm LSR}$\,=\,--30 to +30\,km\,s$^{-1}$) of the H41$\alpha$ recombination line observed
with IRAM (beam size $\sim$30$''$; \citealt{Goicoechea2015}).
Stars show the locations of the Trapezium stars ($\theta^1$ Ori A, B, C,
D, and E). The unit of the colour bars of the middle and right panels is in both cases Kelvin.}
\label{fig:Ratio-Tk-map}
\end{figure*}

The NH$_3$\,(2,2)/(1,1) ratio is sensitive to gas
temperatures $T_{\rm kin}$\,$<$\,50\,K \citep{Mangum2013a,Gong2015a},
which is similar to the kinetic temperature range that the para-H$_2$CO\,(3--2)
ratio is most sensitive to \citep{Mangum1993a}.
The NH$_3$ lines have lower effective excitation densities than
the para-H$_2$CO\,(3--2) transitions by a few orders of magnitude,
$n_{\rm eff}$(NH$_3$(1,1))\,$\sim$\,10$^3$\,cm$^{-1}$ while
$n_{\rm eff}$(para-H$_2$CO\,3$_{03}$--2$_{02}$)\,$\sim$\,10$^5$\,cm$^{-1}$ \citep{Shirley2015}.
The Orion A molecular cloud has been measured in NH$_3$\,(1,1) and (2,2)
with the GBT telescope (beam size $\sim$30$''$; \citealt{Friesen2017};
this agrees well with our para-H$_2$CO\,(3--2) data.)
We compare maps of gas kinetic temperatures derived from
para-H$_2$CO and NH$_3$\,(2,2)/(1,1) line ratios in Figure \ref{fig:Ratio-Tk-map}.
The typical gas kinetic temperature derived from NH$_3$\,(2,2)/(1,1) is 20--30\,K.
Typical gas kinetic temperatures are $>$100\,K in Orion KL, $\sim$50\,K in Orion
South, $>$50\,K in the Orion Bar, 20--30\,K in Orion
north and $>$50\,K in the Orion northeastern region (see Table\,\ref{table:Temperatures}).
Almost everywhere NH$_3$\,(2,2)/(1,1) traces lower kinetic temperatures
than those derived from para-H$_2$CO\,(3--2) line ratios.
We also compare matched pixels of kinetic temperature maps of the two tracers
in Figure\,\ref{fig:Tk-H2CO-NH3}. It seems that temperatures derived from the two
tracers are correlated. In many cases para-H$_2$CO\,(3--2) traces a
higher temperature than NH$_3$\,(2,2)/(1,1) with a difference of 5-->100\,K.
Previous observations toward Galactic CMZ clouds, dense massive
star-forming clumps in our Galactic disk and external galaxies indicate that
in many cases para-H$_2$CO\,(3--2) line ratios
trace a higher kinetic temperature than the NH$_3$\,(2,2)/(1,1) line ratios
\citep{Weiss2001,Muhle2007,Ao2013,Ott2014,Ginsburg2016,Tang2017a,Tang2017b,Tang2017c}.
The difference is likely due to the fact that the derived kinetic temperatures
from NH$_3$\,(2,2)/(1,1) reflect an average temperature
of more extended and cooler gas \citep{Henkel1987,Ginsburg2016} and
para-H$_2$CO\,(3--2) ratios trace denser and hotter regions \citep{Ginsburg2016,Tang2017a,Tang2017c}.

\begin{table*}[t]
\caption{Comparison of temperatures derived from different tracers.}
\centering
\begin{tabular}
{cccccccc}
\hline\hline 

Region &H$_2$CO &NH$_3$ &CH$_3$CCH &CO(1--0) &CO(6--5) &MIR &FIR\\
&K&K&K&K&K&K&K\\
\hline 
KL          &120--290 &>100     &50--60   &78--88 &>150 &>150     &45--60\\
South       &$\sim$74 &$\sim$50 &$\sim$46 &81--92 &>150 &$\sim$75 &$\sim$30\\
North       &$\sim$40 &20--30   &13--40   &43--60 &...  &...      &26--30\\
Northeast   &70--160  &>50      &40--51   &60--78 &...  &...      &30--44\\
Bar (clump) &$\sim$73 &>50      &...      &...    &>150 &$\sim$70 &33--50\\
Bar (edge)  &100--130 &>100     &...      &...    &>100 &>100     &41--51\\
\hline 
\end{tabular}
\tablefoot{Column 2: Kinetic temperatures derived from para-H$_2$CO line ratios
(see Section \ref{sect:Kinetic-temperature}). Column 3: Kinetic temperatures
derived from the NH$_3$\,(2,2)/(1,1) line ratios taken from \cite{Friesen2017}.
Column 4: Rotation temperatures obtained from CH$_3$CCH\,(6--5) taken from \cite{Bergin1994}.
Columns 5 and 6: Excitation temperatures of CO\,(1--0) and CO\,(6--5) taken
from \cite{Bergin1994} and \cite{Peng2012}, respectively. Columns 7 and 8:
Mid (MIR) and far infrared (FIR) dust temperatures taken from \cite{Downes1981},
\cite{Vaillancourt2002}, \cite{Lombardi2014}, \cite{Salgado2016}, and \cite{Kauffmann2017}.}
\label{table:Temperatures}
\end{table*}

A comparison of temperatures obtained
from NH$_3$\,(4,4)/(2,2) and para-H$_2$CO\,(3$_{21}$--2$_{20}$/3$_{03}$--2$_{02}$)
line ratios in a Galactic CMZ cloud indicates that in many cases the tracers
reveal consistent results \citep{Lu2017}. Kinetic temperatures in the Orion Bar have
been measured with the NH$_3$\,(4,4)/(2,2) ratio \citep{Batrla2003},
which shows similar temperatures as derived from the para-H$_2$CO (3--2) ratio.
More highly excited NH$_3$ transitions commonly lead to higher kinetic temperatures
(e.g., \citealt{Henkel1987,Mangum2013a,Gong2015a,Gong2015b}).
Therefore, if higher NH$_3$ levels than NH$_3$\,(2,2)/(1,1) are involved
in measuring the kinetic temperatures, the values derived from
NH$_3$ might become at least as high as those from para-H$_2$CO\,(3--2) in the OMC-1 region.

Several locations in the Orion KL region have been observed in
CH$_3$CCH ($J$\,=\,5--4 and 6--5) \citep{Churchwell1983},
which is also a thermometer tracing dense molecular gas. It reveals a
gas temperature of $\sim$35\,K.
A region similar to that of our observations has been
observed in CH$_3$CCH\,(6--5, beam size $\sim$50$''$; \citealt{Bergin1994}),
suggesting a gas temperature range of 10--60\,K with an average
of $\sim$30\,$\pm$\,1\,K,
which is similar to results obtained from NH$_3$\,(2,2)/(1,1) line ratios
but is lower than that derived from our para-H$_2$CO line ratios.
This difference is also found in dense massive star-forming
clumps \citep{Tang2017c,Giannetti2017}. These indicate that
CH$_3$CCH\,(6--5) might, like the NH$_3$\,(2,2)/(1,1) ratio, trace cooler
and more extended gas \citep{Churchwell1983}.

CO\,(1--0) with a beam size of $\sim$45$''$ has been observed toward the
OMC-1 region \citep{Bergin1994}, which reveals a warm gas
temperature (equivalent to the excitation temperature since $^{12}$CO
is likely optically thick) ranging from 43 to 92\,K with an average
of $\sim$69\,$\pm$\,2\,K.
This is similar to results derived from para-H$_2$CO\,(3--2) line ratios.
Low resolution (beam size $\sim$\,3$'$) observations of the Orion molecular cloud
with CO\,(2--1) show a gas kinetic temperature of $\sim$88\,K
at density $\sim$10$^3$\,cm$^{-3}$ in the Orion KL region \citep{Nishimura2015},
which agrees with our averaged gas temperature ($\sim$82\,K) derived from the
para-H$_2$CO\,(3--2) line ratio at density 10$^5$\,cm$^{-3}$ within a similar region.
Observations of the high excitation CO\,(6--5) transition
(beam size $\sim$\,8.6$''$) toward the OMC-1 region indicate extensive high
gas temperatures (excitation temperature >150\,K) all over Orion KL, Orion South,
and the Orion Bar regions \citep{Peng2012}, whereas in H$_2$CO (and also NH$_3$)
we only see such high $T_{\rm kin}$ values in the Orion KL region.
This suggests that para-H$_2$CO\,(3--2) may trace a cooler component than
the high-excitation transition of CO\,(6--5).

\begin{figure*}[t]
\centering
\includegraphics[width=0.98\textwidth]{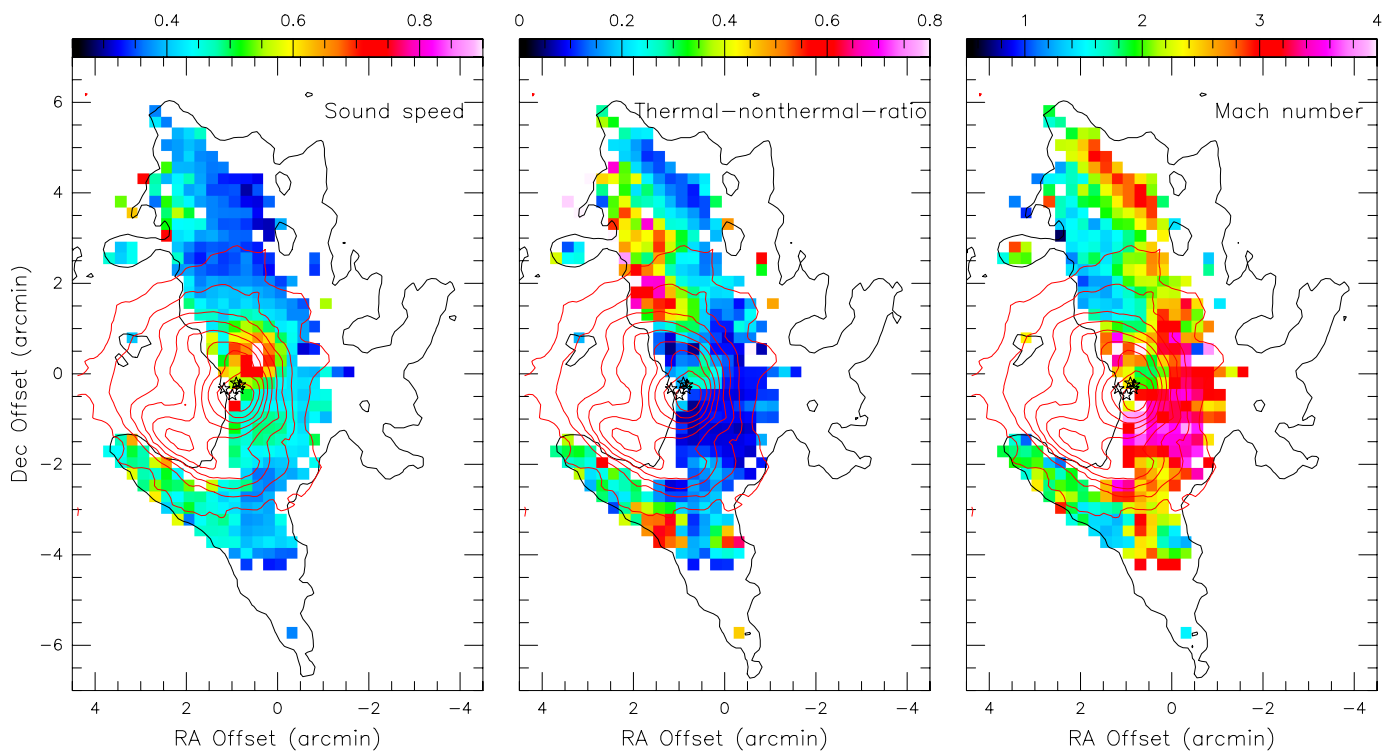}
\caption{Maps of sound speed (left; color bar in units of
km\,s$^{-1}$), thermal to non-thermal gas
pressure (middle), and Mach number (right) in the OMC-1.
Contours and stars are the same as in Figure \ref{fig:Ratio-Tk-map}.
For details, see Section\,\ref{Sec:Thermal and non-thermal motions}}
\label{fig:Thermal-nothermal-ratio-Mach-number}
\end{figure*}

\subsection{Comparison of temperatures derived from the gas and the dust}
\label{sect:dust-temperature}
The dust temperature in the Orion A molecular cloud has been well
studied from far-infrared (FIR) to mid-infrared (MIR) wavelengths
(e.g., \citealt{Downes1981,Mookerjea2000,Vaillancourt2002,
Megeath2012,Lombardi2014,Goicoechea2015,Salgado2016}).
The dust temperatures derived from FIR
measurements rarely exceed 50\,K in star formation regions of
our Galaxy and external galaxies
(e.g., \citealt{Henkel1986,Mangum2013a,Guzman2015,Merello2015,
He2016,Konig2017,Lin2016,Yu2016,Tang2017a}).
Likely, the dust emission at FIR wavelengths originates
primarily from cold dust components that may not be directly
associated with star formation activity
\citep{Schnee2009,Bendo2012,Mangum2013a}.
The cold dust temperatures derived from the FIR measurements have
a range of $\sim$14--60\,K with a roughly constant value, $\sim$30\,K,
in the extended parts of OMC-1 \citep{Mookerjea2000,Vaillancourt2002,
Lombardi2014,Goicoechea2015,Salgado2016},
which agrees remarkably well with results obtained from NH$_3$\,(2,2)/(1,1)
and CH$_3$CCH\,(6--5), but is lower than that derived from
para-H$_2$CO\,(3--2) line ratios. This indicates that the gas
temperatures derived from NH$_3$\,(2,2)/(1,1) and CH$_3$CCH\,(6--5)
tend to be related to the cold dust component responsible for FIR emission.
The dust emission at mid infrared wavelengths traces primarily
warm dust components that may be excited by young stars and
clusters \citep{Helou1986}. The warm dust
temperatures derived from the MIR measurements range from $\sim$45
to $\gtrsim$150\,K in the OMC-1 region \citep{Downes1981,Salgado2016},
which is consistent with results obtained from our para-H$_2$CO line ratios.
This further confirms that the para-H$_2$CO\,(3--2) ratios
trace denser and warmer gas.

At the densest regions ($n$(H$_2$)\,$\gtrsim$\,10$^{4}$\,cm$^{-3}$)
interactions between dust and gas become sufficiently frequent,
so it is commonly expected that gas and dust are thermally coupled \citep{Goldsmith2001}.
Previous observations show that the
temperatures derived from gas and dust are often in agreement
in the active dense clumps of Galactic disk clouds
\citep{Dunham2010,Giannetti2013,Battersby2014}.
Based on measurements of dust at MIR and FIR wavelengths
and gas detected in the para-H$_2$CO\,(3--2), NH$_3$\,(2,2)/(1,1),
and CH$_3$CCH\,(6--5) transitions,
dust and gas temperatures appear to be generally equivalent
in the dense gas ($n$(H$_2$)\,$\gtrsim$\,10$^{4}$\,cm$^{-3}$) of the OMC-1 region,
but yield different values depending on how close they are related to massive star formation.

\subsection{Thermal and non-thermal motions}
\label{Sec:Thermal and non-thermal motions}
Using the kinetic temperatures derived from the para-H$_2$CO line ratios,
the thermal and non-thermal linewidth
($\sigma_{\rm T}$\,=\,$\sqrt{\frac{kT_{\rm kin}}{m_{\rm H_2CO}}}$ and
$\sigma_{\rm NT}$\,=\,$\sqrt{\frac{\Delta v^2}{8{\rm ln}2}-\sigma_{\rm T}^2}$\,$\thickapprox$\,$\Delta v/2.355$,
where $k$ is the Boltzmann constant, $T_{\rm kin}$ is the kinetic
temperature of the gas, $m_{\rm H_2CO}$ is the mass of the formaldehyde
molecule, and $\Delta v$ is the measured FWHM linewidth of para-H$_2$CO
3$_{03}$--2$_{02}$ from a Gaussian fit; \citealt{Pan2009,Dewangan2016})
ranges are 0.09--0.27\,km\,s$^{-1}$ with an average of
0.12\,$\pm$\,0.01\,km\,s$^{-1}$ and 0.34--2.78\,km\,s$^{-1}$ with an average
of 0.98\,$\pm$\,0.02\,km\,s$^{-1}$, respectively. The thermal linewidth is significantly
smaller than the non-thermal linewidth, which indicates that the dense gas
traced by para-H$_2$CO is dominated by non-thermal motions in the OMC-1 region.

Distributions of the sound speed
($a_{\rm s}$\,=\,$\sqrt{\frac{kT_{\rm kin}}{\mu m_{\rm H}}}$,
where $\mu$\,=\,2.37 is the mean molecular weight for molecular
clouds and $m_{\rm H}$ is the mass of the hydrogen atom; \citealt{Dewangan2016}),
the thermal to non-thermal pressure ratio
($R_{\rm p}$\,=\,$a_{\rm s}^2/\sigma_{\rm NT}^2$; \citealt{Lada2003}),
and the Mach number (given as $M$\,=\,$\sigma_{\rm NT}/a_{\rm s}$)
in the OMC-1 are shown in Figure \ref{fig:Thermal-nothermal-ratio-Mach-number}.
The sound speed ranges from 0.30 to 0.97\,km\,s$^{-1}$ with
an average of 0.44\,$\pm$\,0.01\,km\,s$^{-1}$.
The thermal to non-thermal pressure ratio
ranges from 0.05 to 2.12 with an average of 0.27\,$\pm$\,0.01.
The Mach number ranges from 0.7 to 4.3 with an average of 2.3\,$\pm$\,0.1.
All this suggests that non-thermal pressure and supersonic non-thermal
motions (e.g., turbulence, outflows, shocks, and/or magnetic fields)
are dominant in the dense gas traced by para-H$_2$CO in the OMC-1 region.
A few locations have high thermal to non-thermal pressure ratios
($R_{\rm p}$\,$\gtrsim$\,1) corresponding to lower Mach numbers
($M$\,$\lesssim$\,1), which are located in the eastern edge
of the northern part of the OMC-1, in the 10\,km\,s$^{-1}$ filament
(see Figure \ref{fig:Thermal-nothermal-ratio-Mach-number}).

\begin{figure}[t]
\vspace*{0.2mm}
\begin{center}
\includegraphics[width=0.42\textwidth]{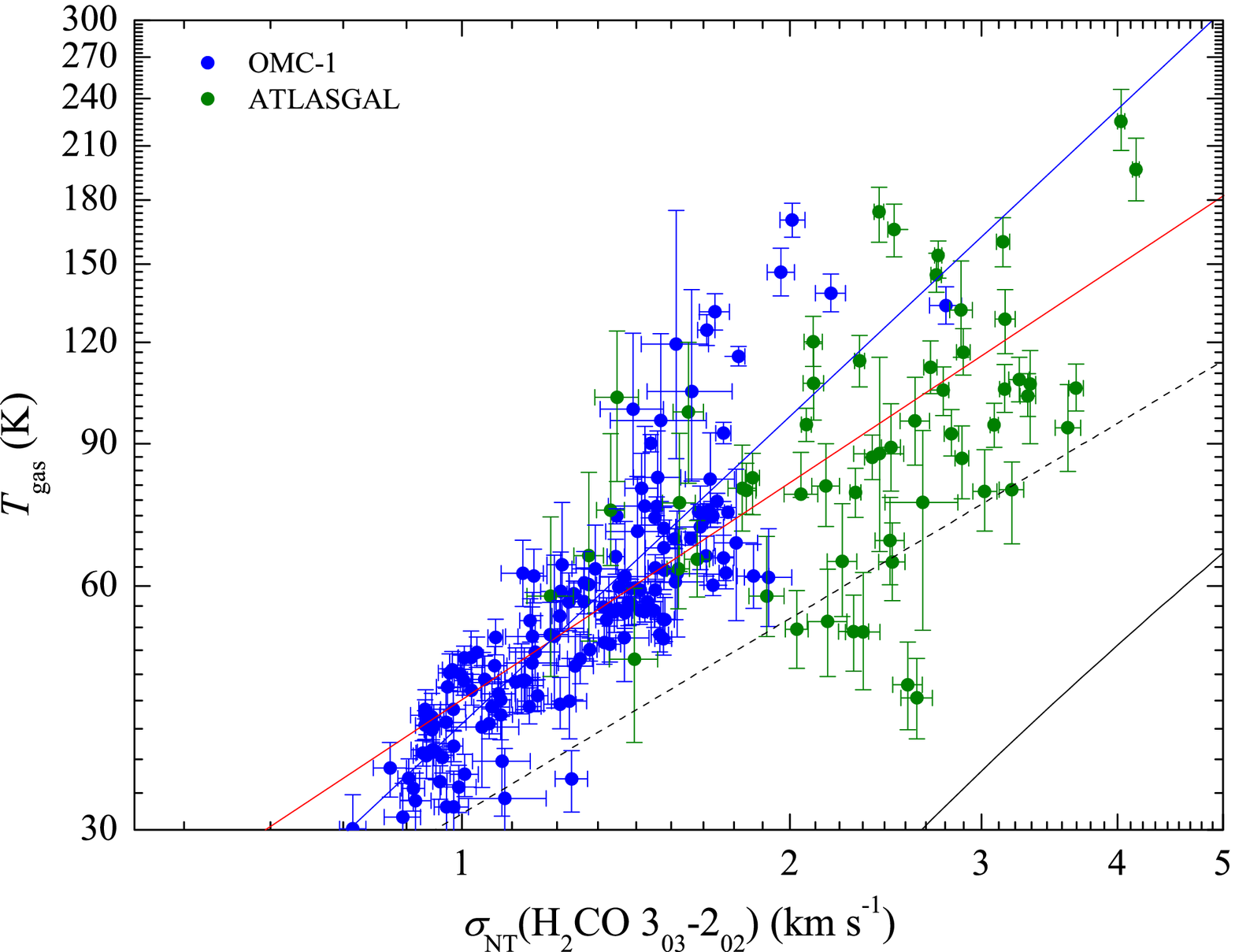}
\includegraphics[width=0.42\textwidth]{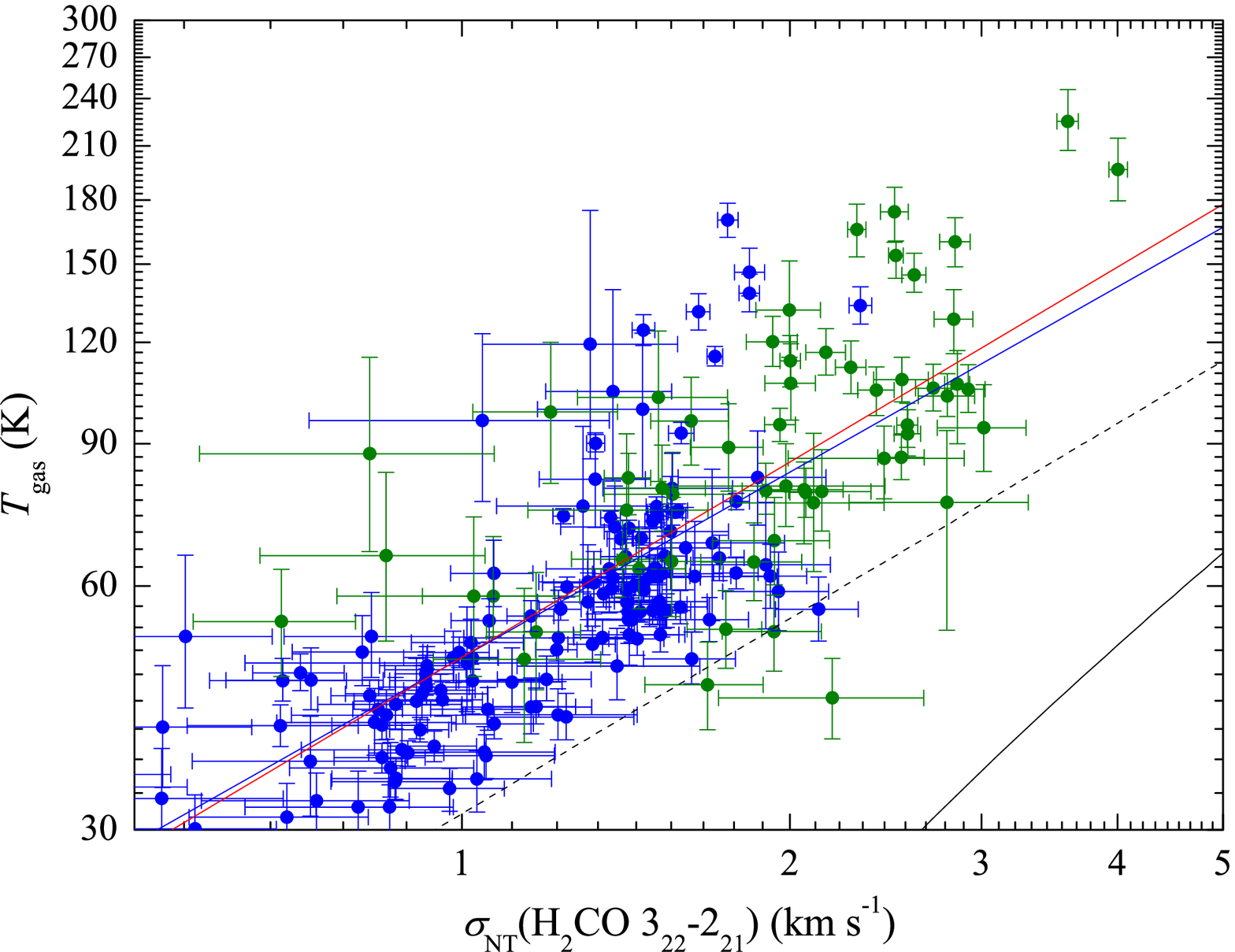}
\includegraphics[width=0.42\textwidth]{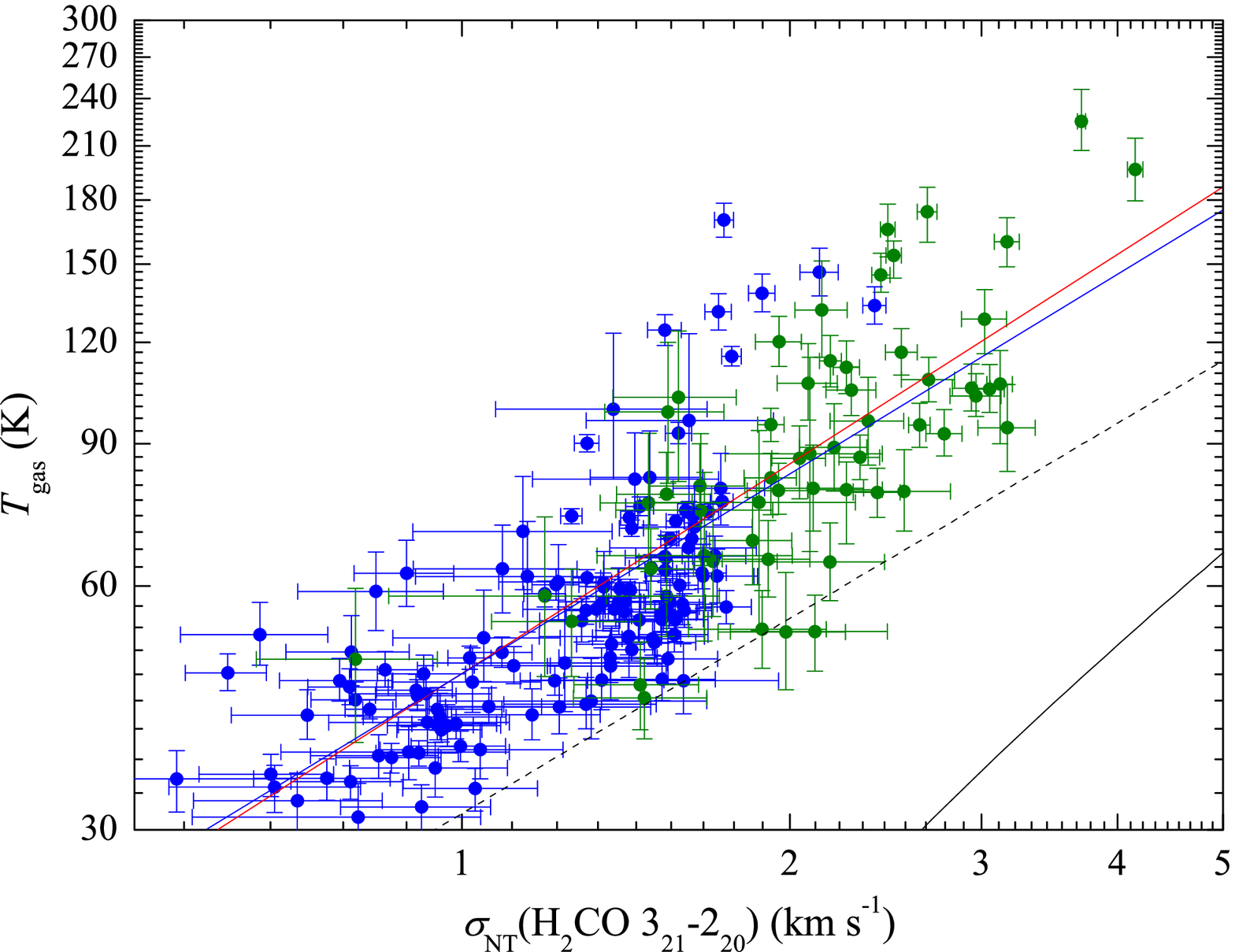}
\end{center}
\caption{Non-thermal velocity dispersion ($\sigma_{\rm NT}$) vs.
gas kinetic temperature derived from the para-H$_2$CO line ratio
(Section \ref{sect:Kinetic-temperature})
for locations with the Mach number $M$\,$\gtrsim$\,2.5 in
the OMC-1 (blue points) and ATLASGAL massive star-forming
clumps (green points; \citealt{Tang2017c}).
The blue line is the result from a linear
fit for the OMC-1 data. The red line is the linearly fitted result
for OMC-1 ($M$\,$\gtrsim$\,2.5) and ATLASGAL data.
Black solid and dashed lines indicate relations between
non-thermal velocity dispersion and
gas kinetic temperature derived from para-H$_2$CO and NH$_3$ at
density 10$^5$\,cm$^{-3}$ in the Galactic CMZ clouds,
respectively \citep{Gusten1985,Immer2016}.}
\label{fig:Tk-width}
\end{figure}

The thermal to non-thermal pressure ratio and Mach number show
apparent gradients along the northern part of the OMC-1
10\,km\,s$^{-1}$ filament and the Orion Bar regions
(see Figure \ref{fig:Thermal-nothermal-ratio-Mach-number}).
This indicates that dense gas probed by para-H$_2$CO may
be affected by the H\,{\scriptsize II} region (see Figures \ref{fig:Ratio-Tk-map} and
\ref{fig:Thermal-nothermal-ratio-Mach-number}).
Positions with higher thermal to non-thermal pressure ratio
($R_{\rm p}$\,>\,0.15) and lower Mach number ($M$\,<\,2.5) located
in the south of the OMC-1 10\,km\,s$^{-1}$ filament and the Orion Bar
region have a higher temperature range
(>60\,K, see Figure \ref{fig:Ratio-Tk-map}) and a lower
linewidth range (<2\,km\,s$^{-1}$, see Figure
\ref{fig:H2CO-velocity-width-distributions}).
This gas may be heated by FUV photons originating from the H\,{\scriptsize II} region.
Low thermal to non-thermal pressure ratios ($R_{\rm p}$\,<\,0.15)
and higher Mach numbers ($M$\,>\,2.5) are associated with the outflows of the
massive star forming regions Orion South and Orion KL, and the northwestern part
of the OMC-1 10\,km\,s$^{-1}$ filament. It indicates that
dense gas traced by para-H$_2$CO is strongly influenced by
non-thermal motions (e.g., outflows and shocks) in these regions.

\subsection{Turbulent heating}
Correlations between the kinetic temperature and linewidth are
expected in the case of conversion of turbulent energy into
heat \citep{Gusten1985,Molinari1996,Ao2013,Ginsburg2016,Immer2016}.
Present observations toward Galactic CMZ clouds with para-H$_2$CO\,(3--2)
show that the warm dense gas is heated most likely by turbulence
on a scale of $\sim$1\,pc \citep{Ao2013,Ginsburg2016,Immer2016}.
However, high-resolution observations
with para-H$_2$CO\,(3$_{21}$--2$_{20}$/3$_{03}$--2$_{02}$)
and NH$_3$\,(4,4)/(2,2) show no apparent correlation between
temperatures and linewidths in a Galactic CMZ cloud
at a smaller scale of $\sim$0.1\,pc \citep{Lu2017}.
Previous observations of, e.g., NH$_3$, H$_2$CO, and CH$_3$CCH in star formation regions
suggest that  the linewidth is correlated with kinetic temperature
\citep{Wouterloot1988,Molinari1996,Jijina1999,Wu2006,
Urquhart2011,Urquhart2015,Wienen2012,Lu2014,Tang2017a,Tang2017c,Giannetti2017}.

We examine whether there is a relationship between
turbulence and temperature on a $\sim$0.06\,pc scale in the OMC-1 region.
We adopt the non-thermal velocity dispersion ($\sigma_{\rm NT}$)
of para-H$_2$CO in good approximation as a proxy for the turbulence,
and the kinetic temperature derived from the para-H$_2$CO line ratio as
the gas kinetic temperature. As mentioned in Section
\ref{Sec:Thermal and non-thermal motions}, the dense gas in locations
with lower Mach number ($M$\,<\,2.5) may be influenced by
external sources, so here we select positions with
strong non-thermal motions ($M$\,$\gtrsim$\,2.5
corresponding to a thermal to non-thermal pressure ratio
$R_{\rm p}$\,$\lesssim$\,0.15), which are located
near massive star formation regions (Orion KL outflows
and Orion South) and the northern region of the
10\,km\,s$^{-1}$ filament of the OMC-1 ridge (see
Figure \ref{fig:Thermal-nothermal-ratio-Mach-number}).
The relation between kinetic temperature and
non-thermal velocity dispersion is shown in Figure \ref{fig:Tk-width}
(blue points; blue fitted lines).
For the non-thermal velocity dispersion of para-H$_2$CO
and kinetic temperature, linear least squares fit results are
listed in Table\,\ref{table:Tk-dv}.
It shows that the non-thermal velocity dispersion of para-H$_2$CO is significantly
positively correlated with the gas kinetic temperatures, especially in the two massive
star formation regions Orion KL and Orion South, by a power-law
of the form $T_{\rm kin}$\,$\propto$\,$\sigma_{\rm NT}^{0.76-1.26}$, which is
consistent with results in massive star-forming clumps
($T_{\rm kin}$\,$\propto$\,$\sigma_{\rm NT}^{0.67-1.06}$, gas kinetic
temperature measured with para-H$_2$CO\,3--2 and 4--3 line ratios, \citealt{Tang2017c}).
This implies that the higher temperature traced by para-H$_2$CO
in the strong non-thermal motion regions of the OMC-1 is related to the
turbulence at a scale of $\sim$0.06\,pc, which clearly
disagrees with at least one cloud from the Galactic center region
on a $\sim$0.1\,pc scale (see \citealt{Lu2017}).

We determine the gas kinetic temperature heated by turbulent
energy following the method applied by \cite{Ao2013}
in their Equation (20),
\begin{eqnarray}
T_{\rm turb} = \Bigg(\frac{(16\times10^{-8}~a^3+7920~a^{0.5}~b~c^3~d^{-1})^{1/2}}{12~b} \nonumber \\
-\frac{4\times10^{-4}~a^{1.5}}{12~b}\Bigg)^{2/3}\,\rm K,
\end{eqnarray}
where the gas density $a$ is in units of cm$^{-3}$,
the velocity gradient $b$ is in units of km\,s$^{-1}$pc$^{-1}$,
the one-dimensional non-thermal velocity dispersion
$c$ is in units of km\,s$^{-1}$, and the
cloud size $d$ is in units of pc.
This leads to a minimum $T_{\rm turb}$ value, because the dust temperature
has been set to $T_{\rm dust}$\,=\,0\,K to simplify the equation \citep{Ao2013}.
We computed the gas kinetic temperature assuming
turbulent heating dominates the heating process. We adopt
a velocity gradient of $b$\,=\,1\,km\,s$^{-1}$pc$^{-1}$
\citep{Bally1987,Wiseman1998,Hacar2017} and
a gas spatial density of 10$^5$\,cm$^{-3}$ (temperature weakly dependent
on the velocity gradient and the gas spatial density; see \citealt{Ao2013}).
We select three locations with the largest linewidths of 2.7, 4.4,
and 6.6\,km\,s$^{-1}$ at Orion north (1.68$'$,4.78$'$) (offsets relative
to our reference position; see Section \ref{sect:Observation}),
Orion South (0.72$'$,--0.72$'$), and Orion KL (0.72$'$,0.48$'$),
corresponding to gas kinetic temperatures derived from the
para-H$_2$CO line ratio of 43, 62, and 133\,K, respectively.
The derived gas kinetic temperature $T_{\rm turb}$ are 24, 58,
and 111\,K at a cloud scale of 0.1\,pc, 9, 22, and 46\,K at
a cloud scale of 0.5\,pc, and 5, 14, and 31\,K at a cloud
scale of 1\,pc in Orion north, Orion South, and Orion KL, respectively.
If the gas temperature derived from para-H$_2$CO is adopted as the dust
temperature (see Section \ref{sect:dust-temperature})
following Equation (7), (8), (18), and (19) in \cite{Ao2013},
Equation (1) has the form
\begin{eqnarray}
3.3\times10^{-27}~a~c^3~d^{-1} = 4\times10^{-33}~a^2~T_{\rm turb}^{1/2}(T_{\rm turb}-T_{\rm dust}) \nonumber \\
+~6\times10^{-29}~a^{1/2}~T_{\rm turb}^3~b
\end{eqnarray}
and the determined gas kinetic temperatures $T_{\rm turb}$ become 52, 87,
and 150\,K at a cloud scale of 0.1\,pc, 42, 62, and 109\,K at
a cloud scale of 0.5\,pc, and 40, 57, and 102\,K at a cloud
scale of 1\,pc in Orion north, Orion South, and Orion KL, respectively.
For the latter two cloud sizes, the obtained $T_{\rm turb}$ values agree reasonably
well with the $T_{\rm kin}$ values derived from our para-H$_2$CO line ratios.
This suggests that turbulent heating significantly contributes
to the gas kinetic temperature on a small scale. However,
these locations are also influenced by e.g., outflows, shocks,
and/or radiation from massive star formation regions in
Orion KL and Orion South \citep{Wiseman1998}.
Nevertheless, turbulent heating may play an important role
to heat the dense gas associated with massive star formation regions
on small scales in the OMC-1.

\begin{table}[t]
\tiny
\caption{Kinetic temperature vs. H$_2$CO non-thermal velocity dispersion.}
\centering
\begin{tabular}
{cccccccccccc}
\hline\hline 
H$_2$CO & Sample & \multicolumn{3}{c}{$T_{\rm kin}$--$\sigma_{\rm NT}$(H$_2$CO)}\\
\cline{3-5}
Transition && Slope & Intercept & $R$ \\
\hline 
3$_{03}$--2$_{02}$ &OMC-1          &1.26(0.06)  &1.61(0.01) &0.85\\
                   &OMC-1+ATLASGAL &0.89(0.05)  &1.64(0.01) &0.81\\
\hline
3$_{22}$--2$_{21}$ &OMC-1          &0.76(0.06)  &1.69(0.01) &0.72\\
                   &OMC-1+ATLASGAL &0.80(0.05)  &1.69(0.01) &0.78\\
\hline
3$_{21}$--2$_{20}$ &OMC-1          &0.82(0.07)  &1.67(0.01) &0.73\\
                   &OMC-1+ATLASGAL &0.86(0.04)  &1.67(0.01) &0.82\\
\hline 
\end{tabular}
\tablefoot{The format of the regression fits is
${\rm log}T_{\rm kin}\,=\,{\rm Slope}\,\times\,{\rm log}\sigma_{\rm NT}(\rm H_2CO)\,+\,{\rm Intercept}$.
$R$ is the correlation coefficient for the linear fit.}
\label{table:Tk-dv}
\end{table}

To check whether there is a relationship between the
turbulence and temperature on scales of 0.06 to 2\,pc,
we combine our data ($M$\,$\gtrsim$\,2.5) with
previous observational results of para-H$_2$CO\,(3--2)
in massive star-forming clumps with a Mach number range of 2.7--6.7
at scale $\sim$0.1--1.8\,pc \citep{Tang2017c}
and fit the relation between the non-thermal velocity
dispersion of para-H$_2$CO and kinetic temperature
in Figure \ref{fig:Tk-width} (red lines).
The least squares linear fit results are
listed in Table\,\ref{table:Tk-dv}.
It shows that the non-thermal velocity dispersion of
para-H$_2$CO is significantly positively correlated with the
gas kinetic temperature by a power-law of the form
$T_{\rm kin}$\,$\propto$\,$\sigma_{\rm NT}^{0.80-0.89}$ at scale from 0.06 to 1.8\,pc,
which is consistent with our results in the OMC-1 region ($M$\,$\gtrsim$\,2.5)
and agrees well with results found with para-H$_2$CO and NH$_3$ in
molecular clouds of the Galactic center
($T_{\rm kin}$\,$\propto$\,$\Delta v^{0.8-1.0}$;
\citealt{Gusten1985,Mauersberger1986,Immer2016}). This indicates that
turbulent heating seems to be widespread in massive star
formation regions on scales of $\sim$0.06--2\,pc.
One should note that this agreement is only in terms of slope,
not of intercept and absolute value (see Figure \ref{fig:Tk-width}).
The fact that the intercept values are different
suggests that there are differences  in how
the gas in the Galactic CMZ clouds and star formation regions of Orion
and ATLASGAL sample are heated.

\subsection{Radiation heating}
\label{Sec:Radiation-heating}
Previous observations of e.g., NH$_3$, H$_2$CO, CH$_3$CN,
CH$_3$CCH, or CH$_3$OH in massive star-forming regions
\citep{Lu2014,Giannetti2017,Tang2017c} suggest
internal radiative heating of embedded infrared
sources upon their surrounding dense gas.
High resolution observations with NH$_3$ (1,1) and (2,2)
toward the OMC-1 show that the dense gas along the northern part of
the OMC-1 10\,km\,s$^{-1}$ filament
is likely heated by radiation from the entire central Orion nebula,
including KL and the Trapezium stars \citep{Wiseman1996,Wiseman1998}.

\begin{figure}[t]
\vspace*{0.2mm}
\begin{center}
\includegraphics[width=0.49\textwidth]{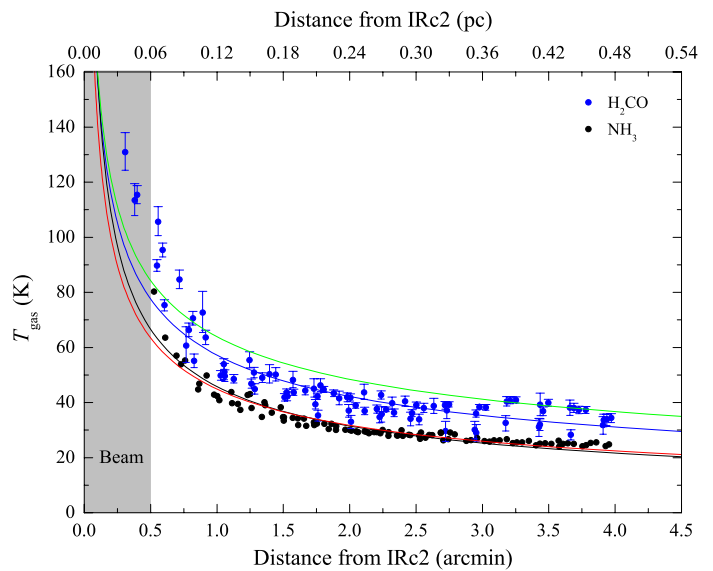}
\end{center}
\caption{The gas kinetic temperature derived from para-H$_2$CO\,(3--2) (blue points)
and NH$_3$ (2,2)/(1,1) (black points) ratios along the northern 10\,km\,s$^{-1}$
filament of the OMC-1 ridge (distance from IRc2,
$\alpha$\,=\,05:35:14.48, $\delta$\,=\,--05:22:30.56, J2000).
The blue and black lines are the fitted results for para-H$_2$CO and NH$_3$,
respectively, with distance $R$\,$>$\,0.5$'$. The red and green lines are the expected
relationships from a Stefan-Boltzmann law and modified Stefan-Boltzmann law
(see Section \ref{Sec:Radiation-heating}), respectively,
assuming Orion KL is the dominant source with an approximate
luminosity of 10$^5$\,${\rm L}_{\odot}$ \citep{Downes1981,Wiseman1998}.}
\label{fig:Tk-distance}
\end{figure}

Following exactly the filament, we investigate the relationship between
gas  kinetic temperature and distance $R$ from the central part of the Orion nebula
(IRc2, $\alpha$\,=\,05:35:14.48, $\delta$\,=\,--05:22:30.56, J2000)
along the northern part of the
OMC-1 10\,km\,s$^{-1}$ filament with para-H$_2$CO and NH$_3$ in Figure \ref{fig:Tk-distance}.
The gas kinetic temperatures are derived from para-H$_2$CO and
NH$_3$\,(2,2)/(1,1) \citep{Friesen2017} line ratios. The beam sizes of para-H$_2$CO
and NH$_3$ are both $\sim$30$''$, so we only fit the data for para-H$_2$CO and NH$_3$,
respectively, with distance $R$\,$>$\,30$''$.
For the gas kinetic temperatures of para-H$_2$CO and NH$_3$,
and distance from IRc2, the fit results are
\begin{equation*}
T_{\rm kin} ({\rm H_2CO}) = (57.2\pm1.1) \times \bigg(\frac{R}{\rm arcmin}\bigg)^{-0.44\pm0.02} ~{\rm and}
\end{equation*}
\begin{equation*}
T_{\rm kin} ({\rm NH_3}) = (45.8\pm0.5) \times \bigg(\frac{R}{\rm arcmin}\bigg)^{-0.54\pm0.01}
\end{equation*}
with power-law indices of --0.44 and --0.54, respectively.
This relation is based on the projected radius along the line of sight.

It is expected that the gas temperature and distance relation from
the Stefan-Boltzmann blackbody radiation law is
$T_{\rm kin}$\,=\,0.86\,$\times$\,$(\frac{L}{{\rm L}_{\odot}})^{1/4}(\frac{R}{\rm pc})^{-1/2}$\,=\,44.8\,$\times$\,$(\frac{R}{\rm arcmin})^{-0.5}$,
adopting a molecular cloud distance of 400\,pc
and assuming that IRc2 is the dominant source with an approximate luminosity
of 10$^5$\,${\rm L}_{\odot}$ \citep{Downes1981,Wiseman1998}, located in the KL nebula.
Adjusting the emissivity of dust grains to be smaller than the
wavelength at the characteristic blackbody temperature,
the radiation law has the form
$T_{\rm kin}$\,$\thickapprox$\,63.9\,$\times$\,$(\frac{R}{\rm arcmin})^{-0.4}$ \citep{Wiseman1998}.
Our fitted power-law indices of para-H$_2$CO and NH$_3$ are remarkably
consistent with the radiation law, which directly
confirms that the dense gas along the northern part of the
OMC-1 10\,km\,s$^{-1}$ filament is heated by radiation from the central Orion nebula.
The two radiation models for gas heating (Stefan-Boltzmann blackbody
radiation and its modification related to dust emissivity)
are both well supported by our para-H$_2$CO and
NH$_3$ data, so here we cannot distinguish which is better.
For the region with $R$\,$\lesssim$\,1$'$,
several locations show gas temperatures probed with
para-H$_2$CO above the fitted results (see Figures \ref{fig:Ratio-Tk-map}
and \ref{fig:Tk-distance}). Indeed, the energy set free by an
explosion resulting from a stellar merger $\sim$\,500 years ago,
is a plausible mechanism causing the high temperatures
of the dense gas in the Orion KL region \citep{BallyZinnecker2005,Bally2017}.

The gas heating appears to be complex and is
most likely due to a number of different processes in the OMC-1 region.
Dense gas around the H\,{\scriptsize II} region at radius $\sim$2.5$'$,
especially the Orion Bar region, appears
to be influenced by FUV photons from the Trapezium stars. Star formation activity
(e.g., outflows, shocks, winds, radiation) in Orion KL and
Orion South also heat the local regions at radius
$\gtrsim$2$'$ \citep{Wiseman1998,Bally2017}.
Turbulent heating seems to be widespread in the dense gas associated
with the massive star formation region in OMC-1.
While dense gas along the northern part of the OMC-1 10\,km\,s$^{-1}$ filament
is heated by radiation from the central Orion nebula.

\section{Summary}
\label{sect:summary}
We have mapped the kinetic temperature distribution of OMC-1 with the
APEX 12\,m telescope, and compared
the kinetic temperatures derived from the para-H$_2$CO 218\,GHz line triplet, with
NH$_3$\,(2,2)/(1,1) inversion lines \citep{Friesen2017} and dust emission.
The main results are the following:
\begin{enumerate}
\item
The distribution of para-H$_2$CO\,3$_{03}$--2$_{02}$  emission is extended
in the OMC-1 region and resembles these of the
NH$_3$\,(1,1) and (2,2) lines on a $\sim$0.06\,pc scale.

\item
The distribution of para-H$_2$CO\,3$_{03}$--2$_{02}$
agrees with the 450 and 850\,$\mu$m dust emission in the OMC-1 region,
suggesting that the H$_2$CO associates well with dense gas traced by the 450
and 850\,$\mu$m continuum.

\item
Using the RADEX non-LTE radiative transfer model, we derive
the gas kinetic temperature modeling the measured
para-H$_2$CO\,0.5$\times$[(3$_{22}$--2$_{21}$\,+\,3$_{21}$--2$_{20})$/3$_{03}$--2$_{02}$]
line ratios. The gas kinetic temperatures derived from para-H$_2$CO
line ratios are warm, ranging from 30 to >200\,K with an
average of 62\,$\pm$\,2\,K at a spatial density of 10$^5$\,cm$^{-3}$.
These temperatures are higher than those obtained from NH$_3$\,(2,2)/(1,1)
and CH$_3$CCH\,(6--5) in the OMC-1 region. In many cases para-H$_2$CO\,(3--2)
traces a higher temperature than NH$_3$\,(2,2)/(1,1) with a difference of 5-->100\,K.

\item
The gas kinetic temperatures derived from the para-H$_2$CO\,(3--2) line ratios agree
with the warm dust components measured at mid-infrared wavelength,
indicating that the para-H$_2$CO\,(3--2) ratios trace
denser and warmer gas than NH$_3$\,(2,2)/(1,1)
and CH$_3$CCH\,(6--5). The cold dust components measured in the far infrared
are consistent with those deduced from
NH$_3$\,(2,2)/(1,1) and CH$_3$CCH\,(6--5). Based on observations
of dust emission at MIR and FIR wavelengths
and molecular gas with para-H$_2$CO\,(3--2), NH$_3$\,(2,2)/(1,1),
and CH$_3$CCH\,(6--5), dust and gas temperatures appear to be
generally equivalent in the dense gas
($n$(H$_2$)\,$\gtrsim$\,10$^{4}$\,cm$^{-3}$) of the OMC-1 region,
but with the MIR continuum and para-H$_2$CO likely sampling gas
more closely associated with massive star formation.

\item
The non-thermal velocity dispersions of para-H$_2$CO are positively
correlated with the gas kinetic temperatures at density
10$^5$\,cm$^{-3}$ in regions of strong non-thermal motion
(Mach number $\gtrsim$\,2.5) in OMC-1, which implies that
the higher temperature traced by para-H$_2$CO is related
to turbulence on a scale of $\sim$0.06\,pc.
Turbulent heating seems to be widespread
in massive star formation regions on a $\sim$0.1--2\,pc scale.

\item
Combining the temperature measurements with para-H$_2$CO\,(3--2) and
NH$_3$\,(2,2)/(1,1) line ratios, we find direct evidence for the
dense gas along the northern part of the
OMC-1 10\,km\,s$^{-1}$ filament being heated by radiation from the central Orion nebula.
\end{enumerate}

\begin{acknowledgements}
The authors are grateful for the valuable comments of the referee.
We thank the staff of the APEX telescope for their assistance in observations.
We also thank Arnaud Belloche and Cosmos Yeh
for their help of observations and data reduction.
The authors are thankful for the helpful comments of Jens Kauffmann.
This work acknowledges support by The Program of the Light in China's Western
Region (LCRW) under grant XBBS201424 and The National Natural Science
Foundation of China under grant 11433008 and 11373062.
This work was partially carried out within the Collaborative
Research Council 956, subproject A6, funded by the
Deutsche Forschungsgemeinschaft (DFG).
C.H. acknowledges support by a Chinese Academy of Sciences President's
International Fellowship Initiative for visiting scientists (2017VMA0005).
Y.G. acknowledges support by The National Natural Science Foundation of China
under grant 11127903 and The National key research and development
program under grant 2017YFA0402702. This research has used NASA's
Astrophysical Data System (ADS).
\end{acknowledgements}

\Online
\begin{appendix} 

\onecolumn
\section{H$_2$CO velocity channel maps}
\begin{figure*}[h]
\vspace*{0.2mm}
\begin{center}
\includegraphics[width=1.0\textwidth]{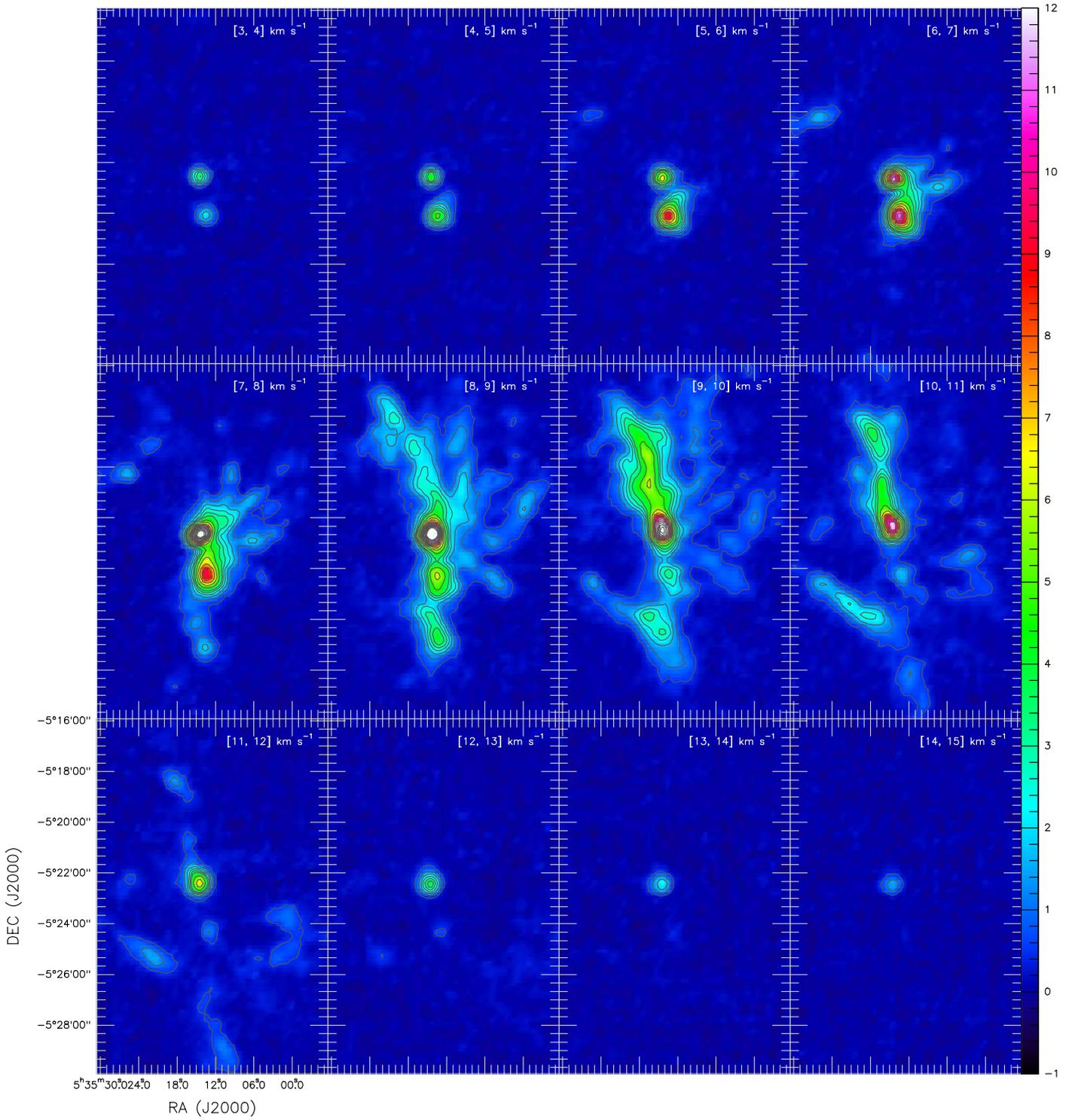}
\end{center}
\caption{Channel maps of the para-H$_2$CO\,3$_{03}$--2$_{02}$ transition.
The contours are running from 0.5 to 2.9 K in steps of 0.6 K and from 3.5
to 14.3 K in steps of 1.2 K ($T$$_{\rm A}^*$ scale; color bar in units of K).}
\label{fig:H2CO-Channel-map}
\end{figure*}

\newpage
\section{Compassion of H$_2$CO, NH$_3$, and dust distributions}
\begin{figure*}[h]
\vspace*{0.2mm}
\begin{center}
\includegraphics[width=1.0\textwidth]{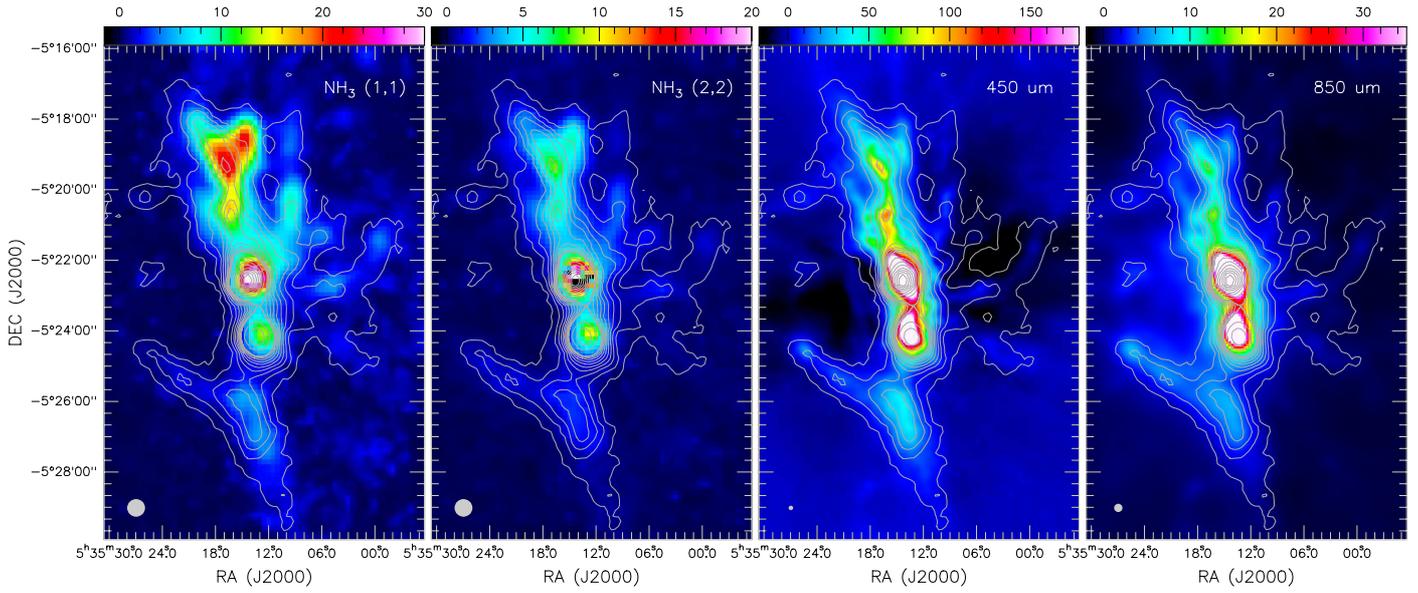}
\end{center}
\caption{Para-H$_2$CO\,(3$_{03}$--2$_{02}$) with
integrated intensity contours (same as in Figure \ref{fig:H2CO-maps}) overlaid on NH$_3$\,(1,1), and (2,2)
integrated intensity observed with the GBT (beam size $\sim$30$''$; left panels; \citealt{Friesen2017}),
and JCMT/SCUBA 450 and 850\,$\mu$m dust emission (beam sizes 7.5$''$ and 14$''$;
right panels; \citealt{Johnston1999}) images of the OMC-1. The beam of
each image is shown in the lower left corner.}
\label{fig:NH3-850um}
\end{figure*}

\newpage
\section{Compassion of para-H$_2$CO\,3$_{22}$--2$_{21}$ and 3$_{21}$--2$_{20}$ line profiles}
\begin{figure*}[h]
\vspace*{0.2mm}
\begin{center}
\includegraphics[width=0.9\textwidth]{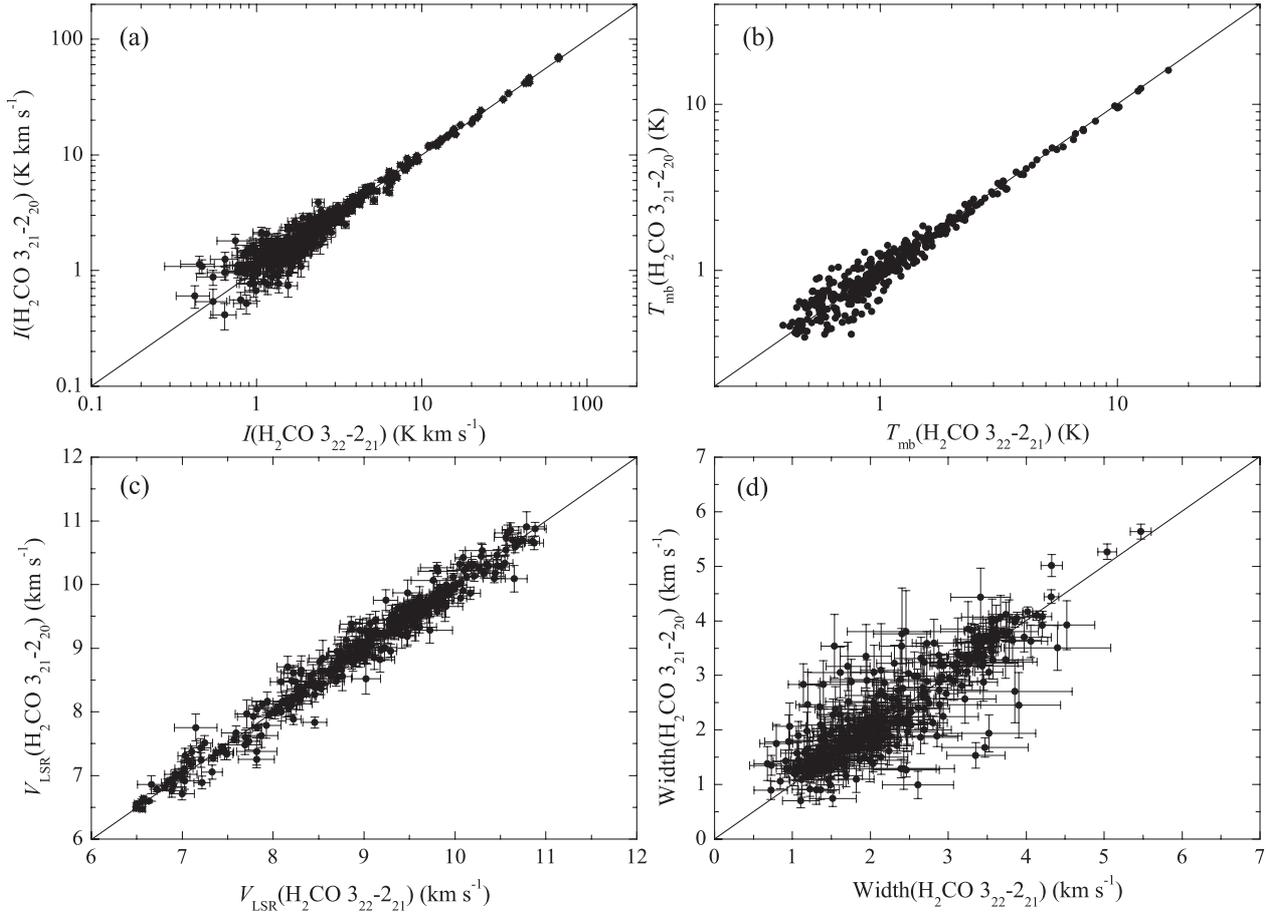}
\end{center}
\caption{Comparisons of integrated intensities (a), line brightness temperatures (b), 
velocities (c), and linewidths (d) of para-H$_2$CO\,3$_{22}$--2$_{21}$ and 
3$_{21}$--2$_{20}$ lines. The straight lines indicate Y\,=\,X.}
\label{fig:H2CO322-321}
\end{figure*}

\newpage
\section{Comparison of gas kinetic temperatures derived from para-H$_2$CO and NH$_3$}
\begin{figure}[h]
\vspace*{0.2mm}
\begin{center}
\includegraphics[width=0.98\textwidth]{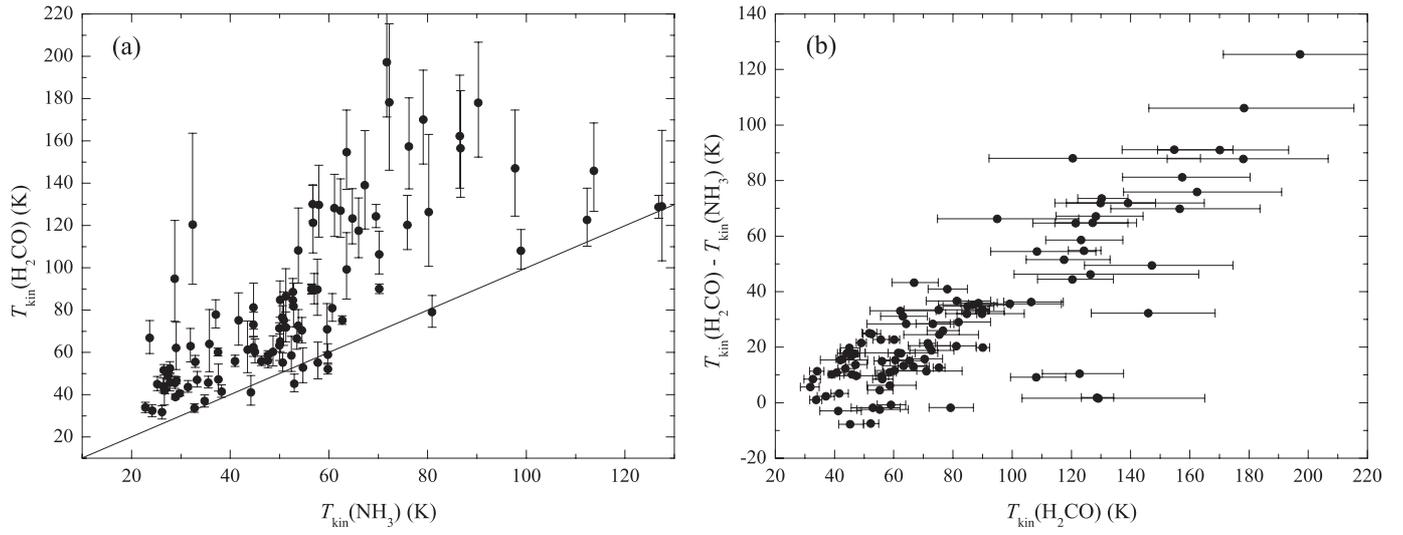}
\end{center}
\caption{Comparison of gas kinetic temperatures
(a) and relative gas kinetic temperature difference (b) derived from
para-H$_2$CO\,(3--2) and NH$_3$\,(2,2)/(1,1) line ratios.
The straight line indicates same temperature.}
\label{fig:Tk-H2CO-NH3}
\end{figure}

\end{appendix}

\end{document}